\newcommand{\beq}{\begin{eqnarray}}
\newcommand{\eeq}{\end{eqnarray}}

\newcommand{\Nf}{N_{\rm f}}

\newcommand{\lqcd}{\Lambda_{\rm QCD}}

\newcommand{\vA}{\vec{A}}
\newcommand{\vB}{\vec{B}}
\newcommand{\vPi}{\vec{\Pi}}
\newcommand{\vcalK}{\vec{\calK}}
\newcommand{\vK}{\vec{K} }
\newcommand{\vp}{\vec{p}}
\newcommand{\vq}{\vec{q}}

\newcommand{\vl}{\vec{l}}
\newcommand{\vL}{\vec{L}}
\newcommand{\vr}{\vec{r}}
\newcommand{\vP}{\vec{P}}

\newcommand{\vR}{\vec{R}}

\newcommand{\la}{\langle}
\newcommand{\ra}{\rangle}

\newcommand{\calL}{\mathcal{L}}

\newcommand{\calB}{\mathcal{B}}

\newcommand{\calN}{\mathcal{N}}

\newcommand{\calK}{\mathcal{K}}

\newcommand{\rmd}{\mathrm{d}}
\newcommand{\rmi}{\mathrm{i}}
\newcommand{\rme}{\mathrm{e}}

\newcommand{\up}{\uparrow}
\newcommand{\down}{\downarrow}

\documentclass[epj]{svjour}
\usepackage{graphics}

\usepackage{verbatim}
\usepackage{bm}
\usepackage{amsmath}
\usepackage{amssymb}
\usepackage{graphicx}
\usepackage{slashed}
\usepackage{wrapfig}
\usepackage{bbm}
\usepackage[caption=false]{subfig}
\usepackage{verbatim}
\usepackage{color}

\begin{document}
\title{Neutral and charged mesons in magnetic fields}
\subtitle{A resonance gas in a non-relativistic quark model }
\author{Toru Kojo\inst{1} }                     
\offprints{}          
\institute{Key Laboratory of Quark and Lepton Physics (MOE) and Institute of Particle Physics, Central China Normal University, Wuhan 430079, China}
\date{Received: date / Revised version: date}
%
\abstract{
We analyze mesons in constant magnetic fields ($B$) within a non-relativistic constituent quark model. 
Our quark model contains a harmonic oscillator type confining potential, 
and we perturbatively treat short range correlations to account for the spin-flavor energy splittings. 
We study both neutral and charged mesons taking into account the internal quark dynamics. 
The neutral states are labelled by two-dimensional momenta for magnetic translations, while the charged states by two discrete indices related to angular momenta. 
For $B \ll \lqcd^2$ ($\lqcd \sim 200$ MeV: the QCD scale), the analyses proceed as in usual quark models, 
while special precautions are needed for strong fields, $B \sim \lqcd^2$, especially when we treat short range correlations such as the Fermi-Breit-Pauli interactions. 
We compute the energy spectra of mesons up to energies of $\sim 2.5$ GeV and use them to construct the meson resonance gas. 
Within the assumption that the constituent quark masses are insensitive to magnetic fields, 
the phase space enhancement for mesons significantly increases the entropy, assisting a transition from a hadron gas to a quark gluon plasma. 
We confront our results with the lattice data, finding reasonable agreement for the low-lying spectra and the entropy density at low temperature less than $\sim 100$ MeV, 
but our results at higher energy scale suffer from artifacts of our confining potential and non-relativistic treatments.
%
\PACS{
      {PACS-key}{discribing text of that key}   \and
      {PACS-key}{discribing text of that key}
     } 
} 
\maketitle

\section{Introduction}

Quantum chromodynamics (QCD) in magnetic fields ($B$) of the QCD scale, $\lqcd \sim 200$ MeV, 
has attracted a lot of attentions in the context of heavy ion physics and the core of neutron stars, and also as laboratories to test theoretical concepts and methodologies  \cite{Miransky:2015ava,Fukushima:2018grm}. 

Lattice Monte-Carlo simulations with magnetic fields do not suffer from the sign problems. 
There have been lattice studies on the chiral and deconfinement transitions \cite{Bali:2011qj,DElia:2018xwo}, various condensates such as chiral condensates \cite{Buividovich:2008wf,Bali:2012zg} and the Polyakov loops \cite{Bruckmann:2013oba}, the string tension \cite{Bonati:2014ksa,Bonati:2016kxj,Bonati:2018uwh}, equations of state \cite{Bali:2014kia}, and hadron spectra \cite{Hidaka:2012mz,Luschevskaya:2018chr,Andreichikov:2016ayj,Luschevskaya:2016epp,Hattori:2019ijy,Bali:2017ian,Ding:2020hxw}. 
The effective model descriptions for these quantities are not straightforward, and the attempts to reproduce the lattice data should improve our understanding of each model.

The lattice studies offer interesting problems concerning the relation between the size of chiral condensates and the chiral restoration temperature. 
Magnetic fields enhance the size of chiral condensates (magnetic catalysis \cite{Klimenko:1991he,Gusynin:1995nb,Suganuma:1990nn}) but reduce the melting (chiral restoration) temperature \cite{Bali:2011qj}. 
The latter is called the inverse magnetic catalysis. 
This phenomenon is in conflict with 
the intuition that a larger chiral condensate leads to a larger dynamical quark mass as well as a larger transition temperature. 
Such intuition is largely based on the experiences from the studies of the Nambu-Jona-Lasinio (NJL) type models with the {\it fixed couplings}; 
at finite $B$, the NJL models lead to significant enhancement in the effective quark masses, chiral condensates, and the restoration temperatures, see, e.g., Refs. \cite{Mizher:2010zb,Gatto:2010pt} 
for early works and recent reviews \cite{Cao:2021rwx,Bandyopadhyay:2020zte}. 
To cure the problem, 
seminal works proposed the $B$-dependent four Fermi couplings in the NJL models \cite{Farias:2014eca,Ferreira:2014kpa,Ferreira:2013tba,Endrodi:2019whh} 
or studied higher order effects such as meson fluctuations or quark-meson couplings \cite{Mao:2016lsr,Mao:2016fha,Ayala:2018zat,Ayala:2020muk}.
We note that, in models without the dynamical quark mass generation, the inverse magnetic catalysis is not necessarily a paradoxical phenomenon;
indeed, for a fixed quark mass, several finite temperature calculations with quarks predict the reduction of the critical temperature, see e.g., Refs.\cite{Fraga:2012fs,Ozaki:2013sfa}.

Other quantities of interest are hadron spectra at finite $eB \gg \lqcd^2$ ($e$: coupling constant in the electrodynamics) which are strong enough to penetrate hadrons and change the internal structure \cite{Fukushima:2012kc,Taya:2014nha}. 
The lattice results differ from the results of hadronic models which neglect the quark substructure. 
The difference is clear-cut for neutral mesons whose mass spectra do not depend on $B$ in hadronic models, but do depend in lattice results \cite{Bali:2017ian}. 
Another example is a charged vector meson whose spin aligns with the magnetic field direction; 
in lattice results the mass at large $B$ tends to be a constant \cite{Bali:2017ian}, while in hadronic models keeps reducing, even leading to the condensations of those mesons \cite{Chernodub:2010qx}. 
For  NJL or quark meson model studies, see, e.g., Refs.\cite{Chernodub:2011mc,Sheng:2020hge,Liu:2018zag,Wang:2017vtn,Avancini:2018svs}.

In the previous works  \cite{Kojo:2012js,Kojo:2013uua,Kojo:2014gha,Hattori:2015aki}, 
we claimed that the dynamically generated quark mass gap at finite $B$ should be $\sim \lqcd$, nearly $B$-independent, and this estimate tames the above-mentioned problems. 
To realize a mass gap of $\sim \lqcd$, it is crucial to examine the range or the momentum dependence of interactions \cite{Braun:2014fua,Mueller:2015fka,Mueller:2014tea,Ayala:2015bgv}. 
For contact interactions without momentum dependence, the solution of the gap equation leads to the mass of $\sim |B|^{1/2}$ which in turn leads to the chiral restoration temperature of $\sim |B|^{1/2}$. 
The $B$-dependence, however, is much milder if long-range interactions (e.g., the $1/r$-type) are used for computations of the quark self-energies,
as interactions with high momentum transfer are much weaker than the contact interactions.
The use of the running $\alpha_s$ further weakens the $B$-dependence in the mass gap. 
The mass gap of $\sim \lqcd$ should lead to the chiral condensate at zero temperature of $\la \bar{q} q \ra \sim |B| \lqcd$, 
the chiral restoration temperature of $T_\chi \sim \lqcd$, 
and the ground state meson spectra approaching constants at large $B$. 
These overall tendencies are in accord with the lattice results. 

In this paper we study the spectra of neutral and charged mesons at finite $B$ within a simple non-relativistic constituent quark model \cite{Zeldovich:1967rt,Sakharov:1980ph,DeRujula:1975qlm,Isgur:1979be}. This model is useful to extract analytic insights which can be readily applied to other models. 

Similar analyses were done for neutral mesons \cite{Simonov:2012if} and fictious\footnote{Mesons should be made of quarks with unequal charges, e.g., $\bar{u}d$ having $-2/3$ and $-1/3$ charges.} charged mesons made of equally charged quarks and antiquarks \cite{Orlovsky:2013gha} for light flavors. There are also studies on light-heavy \cite{Yoshida:2016xgm} and heavy-heavy flavors \cite{Yoshida:2016xgm,Alford:2013jva} for neutral mesons. Mesons at very large $B$ were also analyzed in a relativistic framework but within the lowest Landau level approximation  \cite{Kojo:2012js,Hattori:2015aki}.

Compared to these works, for neutral mesons this work adds some detailed insights on the importance of short-range correlations. 
The treatment of charged mesons is new. In addition to the spectra of mesons at rest, we discuss mesons at finite momenta which are crucial for the estimates of the bulk thermodynamics. 
The phase space enhancement at low energy was originally discussed for neutral pions in Ref.\cite{Fukushima:2012kc} to explain the inverse magnetic catalysis. 
Later Ref.\cite{Hattori:2015aki} (specialized for very large $B$) found the phase space enhancement not only in neutral mesons but also in charged ones, 
reaching the conjecture that such enhancement should assist the chiral restoration as well as the deconfinement. 

In this study we study a wide variety of mesons in the context of a transition from a hadron resonance gas (HRG) to a quark gluon plasma (QGP). 
Such a phase transition takes place through the overlap of hadrons and should accompany the chiral restoration and deconfinement. 
The HRG in magnetic fields were studied in Refs.\cite{Endrodi:2013cs,Fukushima:2016vix} using hadron spectra in the Particle Data Group (PDG) \cite{ParticleDataGroup:2020ssz}
 with the hadronic Zeeman couplings to magnetic fields. 
This approach should be valid for weak magnetic fields $eB \ll \lqcd^2$, but at larger $B$ the structural changes in hadrons should be taken into account.
In this respect the structural changes make some mesons lighter but the others heavier than predicted by the HRG model based on the PDG, 
and hence (except in the large $B$ limit) it is not readily apparent how magnetic fields affect bulk thermodynamic quantities such as pressure and entropy. 
Global analyses of meson spectra are necessary and a simple quark model is suitable for this purpose. 
Including all the above-mentioned effects we find that the entropy is indeed enhanced considerably by magnetic fields, provided that the dynamical quark masses remain $\sim \lqcd$ for a wide range of $B$.

This paper is structured as follows. In Sec.\ref{sec:model} we discuss a quark model and summarize general aspects of the quark dynamics in magnetic fields.
We discuss the spectra of neutral mesons in Sec.\ref{sec:neutral}, 
and charged mesons in Sec.\ref{sec:charged}.
In Sec.\ref{sec:HRG} we discuss the HRG at finite $B$,
comparing the results with the lattice data.
Sec.\ref{sec:discussions} is devoted to discussions. 
We close this paper in Sec.\ref{sec:summary}.

\section{A model and some preparations}
\label{sec:model}

In this section we introduce a model of constituent quarks and summarize methods to be applied for both neutral and charged mesons. 
Our treatment of the quark model is rather standard \cite{DeRujula:1975qlm,Isgur:1979be}, 
but special precautions are given for the short range correlations (given in Sec.\ref{sec:short}) which are crucial for the estimates of the magnetic field effects.

We consider the following hamiltonian in which a quark and an antiquark are moving in constant magnetic fields applied to the $z$-direction, 
\beq 
\hat{H}_0 = \sum_{j=1,2} \bigg[ m_j + \frac{\, \hat{\vPi}_j^2 \,}{\, 2m_j \,} - \hat{ \vec{\mu} }_j \cdot \vB \bigg] + V_{\rm conf} ( \hat{\vr}_1 - \hat{\vr}_2) \,,
\label{eq:hamiltonian}
\eeq
where 
\beq
\hat{ \vPi }_j = \hat{ \vp }_j - e_j\vA_j \,,
\label{eq:covariant_derivative}
\eeq
are kinetic momentum for particles $j=1,2$, and 
\beq
\hat{ \vec{\mu} }_j =  \frac{ e_j }{2m_j } \hat{ \vec{\sigma} }_j \,,
\eeq
are the magnetic moments for which we took the Lande $g$-factor to be 2. 
As a confining potential, we choose a harmonic oscillator
\beq
V_{\rm conf}( \hat{\vr} ) = \alpha \hat{ \vr }^2 \,,
\eeq
which allows us to decompose the hamiltonian into the $z$-dependent and the transverse parts. 
This hamiltonian is regarded as our unperturbed hamiltonian. 
Either magnetic fields or the confining potential make the quark wavefunctions localized. 
After preparing such eigenfunctions, we evaluate the short range effects, such as the Coulomb, color-magnetic interactions, and so on, within a perturbative framework.

\subsection{Conserved quantities}
\label{sec:conserved}

First we find constants of motion.
It is convenient to define pseudo momenta
\beq
\hat{ \vcalK }_j = \hat{ \vPi }_j + e_j (\vB \times \hat{ \vr}_j) \,.
\label{eq:pseudo_mom_def}
\eeq
The kinetic and pseudo momenta satisfy the commutation relations,
\beq
[ \hat{ \Pi}_j^x, \hat{ \Pi}_j^y] =  \rmi e_j B = - [ \hat{ \calK }_j^x, \hat{ \calK }_j^y] \,,~~~~~[ \hat{ \vPi}_j, \hat{ \vcalK}_j ]=0 \,.
\eeq
These commutation relations are valid for any gauge choices.\footnote{
For the pseudo momentum operator, one often starts with the expression $\vcalK' = \vp + e\vA$.
This definition is less general than Eq.(\ref{eq:pseudo_mom_def}). 
The commutation relation $ [ \hat{ \vPi}, \hat{ \vcalK}' ]=0$ is not satisfied for a general gauge choice (except for the symmetric gauge).
} 
In the absence of potentials, pseudo momenta are conserved for each particle, 
but kinetic momenta in the transverse directions are not conserved.
If potentials depend only on $\vr_1-\vr_2$, the sum of pseudo momenta, 
$ \hat{ \vcalK }_R \equiv \hat{ \vcalK }_1+ \hat{ \vcalK }_2$, is conserved,
\beq
[ \hat{H}_0, \hat{ \vcalK }_R ]=0 \,,
\eeq
and the $x$- and $y$-components satisfy the commutation relations,
\beq
[ \hat{ \calK }_R^x, \hat{ \calK }_R^y ] = - \rmi B \sum_{j=1,2}  e_j 
\equiv - \rmi e_R B \,.
\eeq
For charged mesons, only one of the transverse components can be used to label quantum states. 
But for charge neutral mesons ($e_R = 0$),
both $\hat{ \calK }_R^x$ and $\hat{ \calK }_R^y$ are good quantum numbers. 

In addition, the Hamiltonian (\ref{eq:hamiltonian}) can be made axial symmetric around the $z$-axis by choosing the symmetric gauge,
\beq
\vec{A}_j = \frac{\, B \,}{2} (- \hat{y}, \hat{x} )_j = \frac{1}{\, 2 \,} \vB \times \hat{\vr}_j \,,
\eeq
for which
\beq
\hat{ \vPi }_j = \hat{ \vp }_j - \frac{1 }{\, 2 \,} ( \vB_j \times \hat{ \vr}_j ) \,, ~~~
\hat{\vcalK}_j = \hat{\vp}_j + \frac{1 }{\, 2 \,} ( \vB_j \times \hat{\vr}_j ) \,.
\eeq
We wrote $(\vB_j \equiv e_j \vB)$; below we often absorb the charge into $B$ by attaching proper subscripts.
Below our discussions are given in the symmetric gauge.

The orbital angular momentum for the $j$-th particle is $\hat{\vl}_j = \hat{\vr}_j \times \hat{\vp}_j$, and the total orbital angular momentum is
\beq
\hat{\vL} = \sum_{j=1,2} \hat{\vl}_j \,,
\eeq
whose $z$-component commutes with $\hat{H}_0$,
\beq
[ \hat{H}_0 , \hat{L}_z ] = 0 \,.
\eeq
Meanwhile $\hat{L}_z$ does not commute with $\hat{ \calK}_R^{x,y}$; it commutes only with $ \hat{\vcalK}_{R} ^2 $. 
We will see that $ \hat{\vcalK}_{R}^2$ is quantized, and we use $N_{\calK}$ to label the corresponding quantum number.

Therefore we label charge neutral and charged states by quantum numbers
\beq
\big( \calK_R^x \,, \calK_R^y \big)_{\rm neutral} \,,~~~~~ \big( N_\calK \,, L_z \big)_{\rm charged} \,.
\eeq
The former leads to the continuous set of eigenstates while the latter gives the discrete set of integers.

\subsection{Some formulae for the transverse dynamics}
\label{sec:formulae}

\subsubsection{($ \hat{\vPi}^2, \hat{\vcalK}^2$) in polar coordinates}
\label{sec:Pi2}

There are several methods to deal with $\hat{ \vPi }^2$ operators. 
It is convenient to derive relations between the eigenvalues of $ \hat{ \vPi }^2$, $ \hat{ \vcalK }^2$, and $\hat{\vl}_z$,
as they will be used in the perturbative evaluation of short range correlations.
In later sections such relations will be used for several charges,  ($e_j$, $e_R$,...), 
so in the following expressions we will omit the coupling $e$ in front of $B$, and 
will make necessary replacements, $B \rightarrow e_j B$, $B \rightarrow e_R B$, and so on.
We consider the operators (in the symmetric gauge),
\beq
\hat{ \vPi }^2 
&=& \big( \hat{ \vp } - \vA \big)^2 
= \hat{ \vp }^2 + \frac{\, B^2 \,}{4} \hat{ \vr }_\perp^2 - \vB \cdot \hat{ \vl } \,,
\nonumber \\
\hat{ \vcalK }^2 
&=& \big( \hat{ \vp } + \vA \big)^2 
= \hat{ \vp }^2 + \frac{\, B^2 \,}{4} \hat{ \vr }_\perp^2 + \vB \cdot \hat{ \vl } \,,
\label{eq:expand_Pi_calK}
\eeq
where we use the coordinate $\vr = (z, \vr_\perp)$ and $\hat{\vl} = \hat{\vr} \times \hat{\vp}$.
The eigenstates are characterized by ket-vectors $|n_\perp, l\ra$, for which
\beq
\hat{ \vPi }^2/|B| ~&\rightarrow&~  2n_\perp + |l| - l + 1 \equiv 2n_\Pi + 1   \,, \nonumber \\
\hat{ \vcalK }^2/|B|  ~&\rightarrow &~ 2n_\perp + |l| + l + 1  \equiv 2n_\calK+1  \,, 
\label{eq:Pi_calK}
\eeq
or we can invert the relation,
\beq
l =  n_\calK - n_\Pi  \,,~~~~ 
n_\perp = \left\{ \begin{array}{l}
~ n_\Pi~~~~~~\!(l \ge 0) \\
~ n_\calK  ~~~~~~(l < 0) 
\end{array} \right. \,.
\label{eq:l_and_nPi}
\eeq
These expressions will be used to evaluate operators written as functions of $\vr$ and the derivatives.
In coordinate space, we use the polar coordinates $(r_\perp, \theta)$ to express the eigenstate $|n_\perp, l\ra$ as
\beq
\Phi_{B, n_\perp}^l  (\vr_\perp) 
= \sqrt{\frac{\, |B| \,}{2} \,} \tilde{\Phi}_{n_\perp}^l (\vr_B) \,,
		~~~~r_B^2 \equiv \frac{\, |B| \,}{2} r_\perp^2\,,
\eeq
with
\beq
\tilde{\Phi}_{n_\perp}^l (\vr_B) 
	= \calN_\perp \rme^{\rmi l \theta}  r_{B}^{|l|} \rme^{- r_B^2/2} L_{n_\perp}^{ | l | } (r_B^2) \,,
\label{eq:Phi_r}	
\eeq
where $L_{n_\perp}^{ | l | } $ is the associated Laguerre polynomials. With the normalization condition
$\int \rmd^2 \vr_\perp | \Phi (\vr_\perp) |^2 = 1 $, 
 the normalization constant is found to be
\beq
|\calN_\perp |^2 = \frac{1}{\, \pi \,} \frac{\, n_\perp ! \,}{\,  ( n_\perp + |l| )! \,} \,. 
\label{eq:N_perp}
\eeq
%

\subsection{Some formulae for the dynamics in the $z$-direction}
\label{sec:formulae_z}

We need to solve the eigenvalue problem for the relative motion of a particle 1 and 2 in the $z$-direction.
The equation is given by ($z = z_1 - z_2$, $p_z = p_{1z} - p_{2z}$)
\beq 
\bigg( \frac{\, p_z^2 \,}{\, 2\mu \,} + \alpha z^2 \bigg) \psi_{n_z} (z) = E_0^{n_z} \psi_{n_z} (z) \,,
\eeq
where $\mu$ is the reduced mass, $\mu^{-1} = m_1^{-1} + m_2^{-1}$.
The eigenfunction is given by ($H_n$: Hermite polynomials, $n=0,1,2,\cdots$) 
\beq
\psi_{n_z}  (z) = \Lambda_z^{1/2} \tilde{\psi}_{n_z}  (z_\alpha) \,,
~~
z_\alpha = \Lambda_z \, z  \,,
\eeq
where $\Lambda_z \equiv  (2\mu \alpha )^{1/4}$, and
\beq
\tilde{\psi}_{n_z}  (z_\alpha) = \calN_z H_{n_z} (z_\alpha )\, \rme^{- z_\alpha^2/2 }\,,
\eeq
where the normalization constant is
\beq
|\calN_z |^2 = \frac{\, 1 \,}{\, 2^{n_z} n_z! \sqrt{\pi} \,} \,.
\eeq
The eigenvalue is
\beq
E_0^{n_z} = \big(n_z + 1/2 \big) \sqrt{ \frac{\, 2\alpha \,}{\, \mu \,}   \,} \,.
\eeq
With $\mu$ in the denominator,
the energy contribution from the confining effect is larger for lighter quarks.

\subsection{Short range correlations}
\label{sec:short}

Next we consider the short range correlations as perturbations. 
Below we focus on the strong field regime, $|eB| \gg \lqcd^2$, which deserves special considerations. 
(As in Sec.\ref{sec:Pi2}, in this section we omit $e_j$, $e_R$,..., in front of $B$, and later will replace $B$ with $e_j B$, $e_R B$, ..., etc., depending on the situations.) 

Let operators be functions of $\vr^2 = z^2 + \vr_\perp^2$. We evaluate the following types of integrals for operators with the mass dimensions $d$,
\beq
&& \hspace{-0.1cm} 
\la \hat{O}_d (\vr^2) \ra_{n_z, n_\perp}^l \nonumber \\
&&
\equiv \int \rmd z \rmd^2 \vr_\perp \, | \psi_{n_z} (z)|^2 | \Phi_{B, n_\perp}^l (\vr_\perp)|^2 O_d(\vr^2) \nonumber \\
&& 
=
 \int \rmd z_\alpha \rmd^2 \vr_B \, | \tilde{ \psi}_{n_z}  (z_\alpha )|^2 | \tilde{ \Phi}_{n_\perp}^l  (\vr_B)|^2 \tilde{O}_d (z_\alpha, \vr_B ) \,,
\label{eq:exp_value}
\eeq
where
\beq
\tilde{O}_d (z_\alpha, \vr_B ) = \Lambda_z^d \, O_d \bigg( z_\alpha^2 +  \frac{\, \Lambda_z^2 \,}{\, |B| \,} \vr_B^2  \bigg) \,.
\label{eq:O}
\eeq
For low-lying states with $r_B \sim O(1)$, the dependence on $r_B$ apparently drops off at large $B$ and one would expect that the matrix elements are $O(\Lambda_z^d)$. 
But some caution is needed if $O_d$ becomes singular at $\vr \rightarrow 0$ where the details of small $z_\alpha$ become important. 
For example, for the $\delta(\vr)$-type potential with $d=3$, we get
\beq
 \la \delta (\vr) \ra_{n_z, n_\perp}^l  
 &=& | \psi_{n_z} (0)|^2 | \Phi_{B,n_\perp}^l (0)|^2 
 \nonumber \\ 
 &=& \frac{\, | B | \Lambda_z \,}{2} \,  | \tilde{ \psi}_{n_z} (0 )|^2 | \tilde{ \Phi}_{n_\perp}^l  (0)|^2 \,,
 \label{eq:delta-func}
\eeq
which differ from the naive expectation for the large $|B|$ limit of Eq.(\ref{eq:O}). 
For more general analyses, we divide the domain of $z_\alpha$, and examine the leading contributions from $\tilde{O}$. 
Assuming $|\tilde{\psi}_{n_z} |^2 |\tilde{\Phi}_{n_\perp}^l |^2 \sim 1$  for small $z_\alpha$ and $r_B$, 
we estimate the contributions from small $z_\alpha  (< \Lambda_z /\sqrt{ |B| } )$ to be
\beq
\sim \int_0^{ \sim \frac{ \Lambda }{ | B |^{1/2} } r_B } \rmd z_\alpha~ |B|^{\frac{\, d \,}{2} } O (\vr_B^2) 
 \sim
  \Lambda_z | B |^{\frac{\, d-1 \,}{\, 2\,} } O(\vr_B^2) 
 \,.
\eeq
From this expression we see that, for short range interaction with mass dimension $d > 1$, the perturbative corrections becomes very sensitive to the details of $B$. 

For the $d=1$ case ($O\sim 1/r$),  
the logarithms of $B$ arise when we integrate from short $z_\alpha \sim \Lambda_z/\sqrt{|B|}  $ to long distance $z_\alpha \sim 1$, for $r_\calB \sim O(1)$,
\beq
 \int_{\sim \frac{ \Lambda_z }{ | B |^{1/2} } r_\calB }^{\sim 1} \rmd z_\alpha~ z_\alpha^{-1}
 ~ \sim~
  \ln \frac{ \Lambda_z }{ \sqrt{ | B | } } + C \ln r_\calB \,,
\eeq
($C$ is some constant)
as usual logarithmic corrections in perturbation theories.
It is important to stress that, while the logarithm looks weakly dependent on $B$, 
actually the ratio $|B|/\Lambda_z \sim |B|/\lqcd^2$ can be very large for the domain of interest in this paper, 
and the logarithmic $B$-dependence has important impacts on the hadron spectra. 
In QCD, however, such sensitivity to $B$ is largely cancelled if we use the running coupling constant; 
in the current problem the natural renormalization scale should be $|eB|^{1/2}$ and $\alpha_s \sim O(1) /\ln (|eB|/\lqcd^2)$.

Of particular interest in a conventional quark model is the color-electric and color-magnetic interactions at distance scale of $\lesssim 1$ fm,
%
\beq
V_{ E} = - \frac{\, 4 \,}{\, 3 \,} \frac{\, \alpha_s \,}{r} \,,~~~~~~
V_{ M} = V_s (r)\, \vec{\sigma}_1 \cdot \vec{\sigma}_2 \,,
\label{eq:VE_VM}
\eeq
where higher order relativistic corrections are neglected. 

From our scaling analyses we see that $V_E$ corresponds to the $d=1$ case.
Meanwhile, for the spin-spin term in $V_{\rm M}$ we usually use the expression from the Fermi-Breit-Pauli interaction \cite{DeRujula:1975qlm,Isgur:1979be},
\beq
V_{s, {\rm trad} } (r) = \frac{\, \alpha_s \,}{\, m_1 m_2 \,} \, 2\pi \delta (\vr) \,.
\label{eq:Vs}
\eeq
This corresponds to the $d=3$ case leading to the expression Eq.(\ref{eq:delta-func}).
The origin of the delta function is the non-relativistic approximation of the quark-gluon vertex; using the Dirac spinors, the spatial vertex takes the form
\beq
\sim \frac{\, \vec{\sigma}_j \times \vq \,}{\, 2m_j \,} \,,
\eeq
where the strength is proportional to the velocity, and, in the non-relativistic approximation, is proportional to the momentum transfer $\vq$. 
The momenta cancel the $1/\vq^2$ in the gluon propagator so that the product of two vertices and propagators becomes constant in momentum space, leading to the delta function in coordinate space. 
But the magnetic interaction goes back to the expression $\sim 1/r$ when the relativistic effects become important; in this case the velocity becomes $\vq/m \rightarrow \vq/E_q \sim \vq/|\vq|$. 
Therefore the expression (\ref{eq:Vs}) should not be valid for a large $B$ at which the distance between two particles can be very short.

Since the form of the magnetic interactions is sensitive to our non-relativistic approximation, we simply limit its use by introducing a momentum cutoff. 
We use a smooth damping factor $\rme^{-  \vq^2/\Lambda_M^2}$ in momentum space. 
After taking the Fourier transform, in coordinate space it again becomes the Gaussian form,
\beq
V_s (r) =  \frac{\, \alpha_s \,}{\, m_1 m_2 \,} \, C_M \Lambda_M^3 \, \rme^{- \Lambda_M^2 r^2} \,,
\eeq
where the factor $C_M$ will be determined from the hyperfine splitting as in usual quark models \cite{DeRujula:1975qlm,Isgur:1979be}.
In the limit of $\Lambda_M \rightarrow \infty$, the expression goes back to Eq.(\ref{eq:Vs}).
As we have omitted the domain of very large momenta, the expression is regular for small $r$.

\subsection{Choice of parameters}
\label{sec:choice_of_parameters}


\begin{table}[tb]
\caption{Meson masses for $J^{P}=0^{-}$ and $1^-$ states in our constituent quark model. The $l$ refers to $u$- or $d$-quarks. For the experimental values we averaged neutral and charged states. (The mass of $\eta'$ is not shown.)}
\label{tab:meson_mass}       
\begin{tabular}{c lc lc lc |}
\hline\noalign{\smallskip}
  & flavor (theory) & theory [MeV] & experiment [MeV]  \\
\noalign{\smallskip}\hline\noalign{\smallskip}
$\pi$   &~ $\bar{l}l$                                                                           & 140     &~~ 137 \\
$K$    &~ $\bar{l}s$, $\bar{s}l$                                                         & 500     &~~ 497\\
$\eta$ &~ $ \frac{\, \bar{u}u + \bar{d}d - 2 \bar{s}s \,}{\, \sqrt{6} \,}$ & 564     &~~ 547 \\
\noalign{\smallskip}\hline\noalign{\smallskip}
$\rho, \omega$ &~ $\bar{l}l$                                                            & 780      &~~  776 \\
$K^*$ &~ $\bar{l}s$, $\bar{s}l$                                                        & 887      &~~  894 \\
$\phi$ &~ $\bar{s}s$                                                                        & 1014    &~~ 1020 \\
\noalign{\smallskip}\hline
\end{tabular}
\vspace*{0.0cm}  
\end{table}
\begin{table}[tb]
\caption{Impacts of each interaction for various flavor combinations, shown in MeV units.}
\label{tab:interaction}       
\begin{tabular}{c lc lc lc |}
\hline\noalign{\smallskip}
& $m_1+m_2$ & $+\la V_{\rm conf}\ra $   & $+\la V_{E} \ra $  & $+\la V_{\rm s} \ra  $~$(-3\la V_s \ra)$  \\
\hline\noalign{\smallskip}
$ll$   &~~~ 600   &~ 946     &~ 620 &~ 780~  (140)\\
$ls$  &~~~ 800   &~ 1110   &~ 790 &~  887~  (500)\\
$ss$ &~~ 1000 &~ 1268   &~ 955 &\,  1014~ (776)\\
\hline\noalign{\smallskip}
\end{tabular}
\vspace*{0.1cm}  
\end{table}

We choose parameters which reproduce the pseudo-scalar and vector meson masses as done in usual constituent quark models at $B=0$ \cite{DeRujula:1975qlm,Isgur:1979be}.
Including the first order pertrubative corrections from $V_{E,M}$, the meson spectrum takes the form ($M=m_1+m_2$ and $\mu^{-1} = m_1^{-1} + m_2^{-1}$)
\beq
E_{\rm NRQ} (\vK) = \frac{\, \vK^2 \,}{\, 2M\,} + E_{\rm in} \,,
\label{eq:E_NRQ_B0}
\eeq
where $\vK$ is the total momentum, and the meson mass is 
\beq
E_{\rm in} = M + \!\! \sum_{i=x,y,z} \big(n_i +\frac{\, 1 \,}{2} \big)  \sqrt{ \frac{\, 2\alpha \,}{\, \mu \,}   \,} + \la V_E \ra + \la V_M \ra_S \,,
\label{eq:meson_B0}
\eeq
where $\la V_M \ra_{S} = \la V_s \ra \big( 2S(S+1)-3 \big)$ for the total spin $S$.

We fit our low lying meson spectra to the experimental values.
We choose
\beq
&& \alpha = (0.160\, {\rm GeV})^3 \,, \nonumber \\
&& m_{ud} = 0.30\, {\rm GeV}\,,~~~~~~ m_s = 0.50\, {\rm GeV}\,, \nonumber \\
&& \Lambda_M = 1.0\, {\rm GeV}\,,~~~~~~ C_M = 2.02 \,.
\eeq
These parameters are correlated through our fit. 
A larger $\alpha$ demands a larger strength of $V_E$.
We try not to use the parameter set for which $\la V_E \ra \lesssim -400$ MeV (see below).

 For $\alpha_s$, we use the running coupling parameterized as
\beq
\alpha_s (Q) = \frac{\, 4\pi \,}{\, \beta_0 \ln \frac{\, Q^2 \,}{\, \Lambda_{\rm NP}^2 \,} \,} 
 \,.
\eeq
Here $\beta_0 = 11- 2\Nf/3$ for which we simply take the three flavor value $\Nf=3$ for all $Q^2$, 
and the nonperturbative renormalization scale $\Lambda_{\rm NP}$ is $\sim 0.2-0.3$ GeV whose precise value depends on the renormalization scheme
and higher order loops.
Although $\alpha_s$ at energy $\lesssim 1$ GeV contains several uncertainties related to the nonperturbative effects, 
we will use the present running form just for a theoretical orientation.
We take $\Lambda_{\rm NP} \simeq 0.25$ GeV with which $\alpha_s(1\,{\rm GeV}) \simeq 0.5$, reasonably consistent with Fig.3.1 in Ref.\cite{Deur:2016tte}.
When this $\alpha_s(Q)$ is used for our quark model calculations,
we set
\beq
 Q^2 = \Lambda_z^2 + Q_\perp^2 /2\,,
 \label{eq:Q2}
\eeq
where $Q^2_\perp \sim B$, and its precise form will be given when we discuss neutral and charged mesons, see Secs.\ref{sec:neutral_pert} and \ref{sec:charged_pert}.
This choice interpolates the weak and strong field regimes;
for a small $B$, the typical momentum transfer is characterized by the size of hadrons, $\sim \Lambda_z^{-1}$, determined by the confining scale,
while at large $B$ the relevant size scale is $\sim B^{-1/2}$. 
The factor $1/2$ reflects that for a  momentum transfer we choose one direction and $B \sim \langle \hat{p}_x^2 \rangle + \langle \hat{p}_y^2 \rangle$.
With these parameters, the masses of ground state pseudo-scalar and vector mesons are reproduced well (Table.\ref{tab:meson_mass}).

It is important to know the energy budget of each interaction as they react to $B$ differently. 
Shown in Table.\ref{tab:interaction} are the meson masses with successive addition of the potential energy where the sizes of wavefunctions play important roles. 
The confining potential leads to the zero point energy of 300-400 MeV, larger for lighter quarks. This energy cost is largely cancelled by the color-electric interactions of $-$(300-400) MeV, where the impact is larger for heavier quarks as they are more compactly localized. Finally the color-magnetic potentials $\la V_s\ra$ are of 60-150 MeV which are larger for lighter quarks. The last two short-range correlation effects become more important for a larger $B$, as we will see later.



\section{Neutral mesons}
\label{sec:neutral}

We first discuss neutral mesons. 
We begin to prepare unperturbed bases which solve the harmonic oscillator problem in a magnetic field. 
In the next step we consider various perturbations, especially those related to the mass differences and short range correlations.

\subsection{Unperturbed bases}
\label{sec:neutral_unpert}

For neutral mesons, $\vcalK_R$ is conserved in all directions so that it is convenient to begin with the eigenstates for these operators. 
Below we write $q = e_1 = - e_2$ and write $\vB_q = q \vB$. 
Choosing the center of mass coordinates,
\beq
\vR = \frac{\, m_1 \vr_1 + m_2 \vr_2 \,}{M} \,,~~~~~\vr = \vr_1 - \vr_2 \,,
\eeq
where $M=m_1 +m_2$, then the pseudo momentum becomes
\beq
\hat{\vcalK}_R = - \rmi \frac{\partial }{\, \partial \vR \,} + \frac{1 }{\, 2 \,} ( \vB_q \times \hat{\vr} ) \,.
\eeq
For a given $\vr$, solutions $\Phi_{\vK}$ for the eigenvalue problem $\hat{\vcalK}_R \Phi_{\vK} = \vK \Phi_{\vK}$ take the form,
\beq
\Phi_{\vK} (\vR, \vr) = \exp\bigg[ \rmi \vR \cdot \bigg( \vK - \frac{1 }{\, 2 \,} \vB_q \times \vr \bigg) \bigg] \,.
\eeq
For a given eigenvalue $\vK$, the eigenfunction of the hamiltonian can be written as
\beq
\Psi_{\vK} (\vR, \vr) = \Phi_{\vK} (\vR, \vr) \varphi_{\vK} (\vr) \,.
\eeq
It is clear that the center of motion and relative motion couple.
For this form of wavefunctions, our eigenvalue problem, 
$\hat{H}_0 \Psi = E_0 \Psi$, 
can be reduced to $\big( \Phi^* \hat{H}_0 \Phi \big) \varphi = \hat{H}_0' \varphi = E_0 \varphi$, 
where\footnote{We found that our sign of the $\vB_q \cdot \hat{\vl}$ term is in conflict with Refs.\cite{Yoshida:2016xgm} and \cite{Alford:2013jva}. 
 }
(reminder: $\mu = m_1 m_2/M$)
\beq
\hat{H}_0'
&=& M+ \frac{\, \vK^2 \,}{\, 2M \,}
+ \frac{\, \hat{\vp}^2 \,}{\, 2\mu \,} + \alpha \hat{\vr}^2 -  \hat{ \vec{\mu} } \cdot \vB
\nonumber \\
&&+\, \frac{\, B_q^2 \,}{\, 8\mu \,} \hat{\vr}_\perp^2 - \frac{1}{\, M \,} \vB_q \cdot ( \hat{\vr} \times \vK ) 
- g_{\Delta m}  \vB_q \cdot \hat{\vl}
\,.
\eeq
Here, $\hat{\vp} = -\rmi \partial/\partial \vr$, $\hat{\vl} = \hat{\vr} \times \hat{\vp}$, $\hat{ \vec{\mu} } =\hat{ \vec{\mu} }_1 + \hat{\vec{\mu} }_2$, 
and 
\beq
g_{\Delta m} \equiv  \frac{\, 1 \,}{2} \bigg( \frac{1}{\, m_1 \,} - \frac{1}{\, m_2 \,} \bigg) \,.
\label{eq:def_calB}
\eeq
Choosing the eigenstates for the dynamics in the $z$-direction, we now have
(reminder: $E_0^{n_z} = \big(n_z + 1/2 \big) \sqrt{ 2\alpha / \mu   \,} $)
\beq
\hat{H}_0'
&=& M+ \frac{\, \vK^2 \,}{\, 2M \,}
+ E_0^{n_z}
- \hat{\vec{\mu} } \cdot \vB
+ \hat{H}^\perp
\,,
\eeq
where
\beq
\hat{H}^\perp
&=& 
 \frac{\, \hat{\vp}_\perp^2 \,}{\, 2\mu \,} 
 +\, \frac{\, \calB_q^2 \,}{\, 8\mu \,} \hat{\vr}_\perp^2
- \frac{1}{\, M \,} \vB_q \cdot ( \hat{\vr} \times \vK ) 
- g_{\Delta m} \vB_q \cdot \hat{\vl}
\,, \nonumber\\
\eeq
with
\beq
\calB_q^2  \equiv B_q^2 + 8 \mu \alpha \,.
\label{eq:def_calB}
\eeq
For $\hat{H}^\perp$, we eliminate terms linear in coordinates ($\propto \hat{\vr} \times \vK$) by shifting the coordinates. We first introduce the parameter
\beq
\eta 
= 
\frac{\, 4 \mu \,}{M} \bigg( \frac{\, B_q \,}{\, \calB_q \,} \bigg)^2 ~~(\le 1)\,.
\eeq
for later convenience, and make a shift
\beq
\hat{\vr}  \rightarrow \hat{\vr} -  \eta \vec{e}_z \times \frac{\, \vK \,}{B_q}\,.
\eeq
Then $\hat{H}^\perp = \hat{H}_0^\perp + \delta \hat{H}_0^\perp$ is
\beq
\hat{H}_0^\perp 
=  \frac{\,1 \,}{\, 2\mu \,} \bigg( \hat{\vp}_\perp^2 + \frac{\, \calB_q^2 \,}{4} \hat{\vr}_\perp^2 \bigg)  
- \eta  \frac{\,  \vK_\perp^2  \,}{\, 2M \,} 
\,,
\label{H0_perp}
\eeq
and
\beq
\delta \hat{H}_0^\perp 
= - g_{\Delta m}  \bigg( \vB_q  \cdot \hat{\vl} + \eta \vK_\perp \cdot \hat{\vp}_\perp \bigg)\,.
\label{eq:H_0perp}
\eeq
The second term in $\delta \hat{H}_0^\perp$ comes from
the shift of $\hat{\vr}_\perp $ in the $\hat{\vr}_\perp^2$ term. 
The $\delta \hat{H}_0^\perp$ is non-vanishing only for $m_1\neq m_2$, and is treated as a perturbation.

In Sec.\ref{sec:formulae} we have seen  how to deal with the 2D harmonic oscillator, $\hat{H}^\perp_0 |n_\perp, l \ra = E_0^\perp | n_\perp, l \ra $.
We use Eq.(\ref{eq:Pi_calK}) with replacement $B \rightarrow \calB_q$ to derive
\beq
E_0^\perp (n_\perp, l) = \frac{\, |\calB_q| \,}{\, 2 \mu \,} \big( 2n_\perp + |l| + 1 \big) - \eta  \frac{\,  \vK_\perp^2  \,}{\, 2M \,}  \,.
\eeq
Combining all these pieces, for the bases $| \vK; n_\perp, l, n_z \ra$, our hamiltonian can be written as
\beq
\hat{H}_0' = E_K + E_0^{\rm rel} + \hat{H}_0^{\rm spin} + \delta \hat{H}_0^\perp \,,
\eeq
where
\beq
E_K &=& \frac{\, K_z^2 + (1-\eta) \vK_\perp^2 \,}{2M} \,, \nonumber \\
E^{\rm rel}_0 
&=& M 
+ E_0^{n_z}
+ \frac{\,| \calB_q| \,}{\, 2 \mu \,} \big( 2n_\perp + |l| + 1 \big)
\,, \nonumber \\
\hat{H}_0^{\rm spin} 
&=& 
- \hat{ \vec{\mu} } \cdot \vB
 \,.
 \label{eq:E_K}
\eeq
We note that, at large $B$, the factor $(1-\eta) < 1$ suppresses the $\vK_\perp^2$ kinetic term\footnote{In particular for equal masses, $4\mu/M=1$, and at large $B$,
\beq
1-\eta =  1- B_q^2/\calB_q^2 ~\sim~ 8\mu \alpha/B_q^2 ~\sim~ \lqcd^4/B_q^2 \,,
\eeq
so that the coefficient of $\vK_\perp^2$ are suppressed. Also $\delta \hat{H}_\perp=0$ in this case.}.

\subsection{Perturbations}
\label{sec:neutral_pert}

We examine various perturbations using the unperturbed bases, $\Psi (\vr) = \psi(z) \Phi (\vr_\perp)$, in the last section.
Our perturbative hamiltonian is (see Eqs.(\ref{eq:VE_VM}) and (\ref{eq:H_0perp}))
\beq
\hat{H}_1 = \delta \hat{H}_0^\perp + V_E + V_M \,,
\eeq
for which we apply the first order perturbation theory.

The first order perturbation from $\delta H_0^\perp$ is simple. It depends only on the angular momentum $ l$, so we write
\beq
\la \delta \hat{H}_0^\perp \ra_{l}
=  - g_{\Delta m}  l B_q 
= - \frac{\, B_q l \,}{2} \bigg( \frac{1}{\, m_1 \,} - \frac{1}{\, m_2 \,} \bigg) \,.
\eeq
Here we have used $\la \hat{\vp}_\perp \ra = 0$ because the operator $\hat{\vp}_\perp$ raises or reduces the orbital Landau level by one. 
The effects of the $\hat{\vp}_\perp$ term appear from the second order perturbation\footnote{The second order effects 
need the excitation energy of $\sim |\calB_q|/\mu$, and the hopping matrix elements are $\sim \sqrt{ |\calB_q| }$, so the second order correction to the energy is
\beq
\sim \eta^2 g_{\Delta m}^2 \mu \vK_\perp^2 \,.
\eeq
The $g_{\Delta m} =0$ for $m_1=m_2$.
If we take $m_1 = m_{u,d}$ and $m_2 = m_s \simeq 5 /3 \times m_{u,d}$, then $\mu = 5m_u/8$ and $g_{\Delta m} =  2/5m_u$, 
and this second order correction is numerically suppressed.
In the following we will ignore the second order corrections.
}.

The first order perturbative corrections from short range correlations $\la V_E \ra$ and $\la V_M \ra$ were discussed in Sec.\ref{sec:short}. 
For $Q_\perp^2$ in $\alpha_s(Q^2)$ (see Eq.(\ref{eq:Q2})), we use $Q_\perp^2 = \calB_q$.
Using the unperturbed bases $| \vK; n_\perp, l, n_z \ra$ to take the expectation values of orbital wavefunctions, we estimate the hamiltonian as
(see Eq.(\ref{eq:E_K}))
 \beq
 \la \big( \hat{H}_0' + \hat{H}_1 \big) \ra_{\rm orbital}
 = E_K + \hat{H}_{\rm in} \,,
 \label{eq:E_in}
 \eeq
 where $\hat{H}_{\rm in} \equiv E_{0+1}^{\rm rel} +  \hat{H}_{0+1}^{\rm spin} + \la \delta \hat{H}_0^\perp \ra$ with
 (``$0+1$" indicates the sum of the 0th and 1st order hamiltonians)
\beq
E_{0+1}^{\rm rel}
&=& E_{0}^{\rm rel} + \la V_E \ra \,,  
	\nonumber \\
\hat{H}_{0+1}^{\rm spin}
&=&  - \hat{ \vec{\mu} }  \! \cdot \!\vB + \la V_s \ra \,  \hat{ \vec{\sigma} }_1 \cdot \hat{ \vec{\sigma} }_2  \,.
\label{eq:H_spin}
\eeq
Below we find the spin eigenstates for $\hat{H}_{0+1}^{\rm spin}$.
The eigenvalue of $\hat{H}_{\rm in}$ is written as $E_{\rm in}$.

\subsection{Spin dependent terms}
\label{sec:spin_dep_terms}

Now we examine the Zeeman splitting term $\hat{H}_{\rm spin}^0$,
\beq
- \hat{ \vec{\mu} } \cdot \vB
=  \frac{\, B_q \,}{2}  
\times
\left\{ \begin{array}{l}
- \big(\frac{1}{m_1} - \frac{1}{m_2}\big) = - \frac{\, \Delta m_{21} \,}{\, \mu M \,}~~\!(\up \up) \\
+ \big(\frac{1}{m_1} - \frac{1}{m_2}\big) = + \frac{\, \Delta m_{21} \,}{\, \mu M \,}~\,(\down \down) \\
+ \big(\frac{1}{m_1} + \frac{1}{m_2}\big) = -\frac{1}{\, \mu \,}~~~~~~\,(\down \up) \\
- \big(\frac{1}{m_1} + \frac{1}{m_2}\big)= +\frac{1}{\, \mu \,}~~~~~~\,(\up \down) 
\end{array} \right.
\eeq
Assuming $q>0$, the largest energy reduction is achieved for the $(\up \down )$ combination in which both particles (with $q = e_1 = - e_2$) 
can occupy the lowest Landau level, and this energy reduction tends to cancel the zero point energy from the transverse kinetic terms;
\beq
 \frac{\, |\calB_q| \,}{\, 2 \mu \,} - \bigg(\frac{1}{m_1} + \frac{1}{m_2}\bigg) \frac{\, B_q \,}{2} 
 =  \frac{\, |\calB_q| - B_q \,}{\, 2 \mu \,} ~\sim~ \frac{\, 4\mu \alpha \,}{B_q} \,,
 \label{eq:cancellation}
\eeq
for a large $B$. 
So the ground state energy of neutral mesons made of $(\up \down)$ asymptotically becomes insensitive to $B$. 
The other states have the energies of $O(B/\mu)$.

The $\hat{ \vec{\sigma} }_1 \cdot \hat{ \vec{\sigma} }_2$ term in Eq.(\ref{eq:H_spin}) prefers the total spin bases $|S, S_z\ra $ 
and is dominant at weak $B$; this term is responsible for the energy splitting between e.g., $\pi_0$- and $\rho_0^{S_z=0}$-mesons.
Such distinction becomes blurred for a large $B$ where the $ \hat{ \vec{\mu} }  \! \cdot \!\vB$ term is dominant. 
Here the $|s_z\ra_1 \otimes |s_z\ra_2$ bases, which reflect the Landau level structure of particles 1 and 2, become more appropriate.

To cover both small and large $B$ domains, we first write the matrix elements of $\hat{H}_{0+1}^{\rm spin}$ for the bases 
($|S=1,S_z=1\ra, |S=1, S_z=-1\ra, |S=1, S_z = 0\ra, |S=0, S_z = 0 \ra$),
and then diagonalize it.
 The spin-aligned components are diagonal from the beginning,
\beq
\big(E^{\rm spin}_{0+1} \big)_{S=1, S_z=\pm 1} 
= \la V_s \ra \mp \frac{\, B_q \,}{\, 2\mu \,}  \frac{\, \Delta m_{21} \,}{\, M \,}\,,
\eeq
while for the $S_z=0$ components we diagonalize the matrix
\beq
\left[
\begin{matrix}
~~  \la V_s \ra &~~   \frac{\, B_q \,}{\, 2\mu \,}   \\
~~   \frac{\, B_q \,}{\, 2\mu \,}  ~&~ -3 \la V_s \ra ~\\
\end{matrix} \right]
\nonumber 
\eeq
which leads to the eigenvalues for the $|S_z= 0^\pm \ra$ states,
\beq
\big(E^{\rm spin}_{0+1} \big)_{S_z= 0^\pm } = - \la V_s \ra  \pm \sqrt{ 4 \la V_s \ra^2 + (B_q/2\mu)^2 \,} \,.
\eeq
For small $|B_q|$,
\beq
\big(E^{\rm spin}_{0+1} \big)_{S_z= 0^\pm} 
	&=& \la V_s \ra + B_q^2/8\mu^2 \la V_s \ra + \cdots \nonumber \\
\big(E^{\rm spin}_{0+1} \big)_{S_z= 0^\pm } 
	&=& -3 \la V_s \ra - B_q^2/8\mu^2 \la V_s \ra + \cdots
\eeq
and for large $B$,
\beq
\big(E^{\rm spin}_{0+1} \big)_{S_z= 0^\pm}
	= \pm |B_q| /2 \mu - \la V_s \ra + \cdots \,.
\eeq
As we have mentioned in Eq.(\ref{eq:cancellation}), 
for the $S_z = 0^-$ state the zero point energy (the $+ |\calB_q|/2\mu$ term in $E_{0+1}^{\rm rel}$) 
and the Zeeman term (the $- |B_q|/2\mu$ term in $\big(E^{\rm spin}_{0+1} \big)_{S_z=0^- } $) largely cancel. 
In particular, the states with $n_\perp = l =0$ and $|S_z=0^- \ra$ has the energy weakly dependent on $B$.

\begin{figure}
\resizebox{0.5\textwidth}{!}{%
  \includegraphics{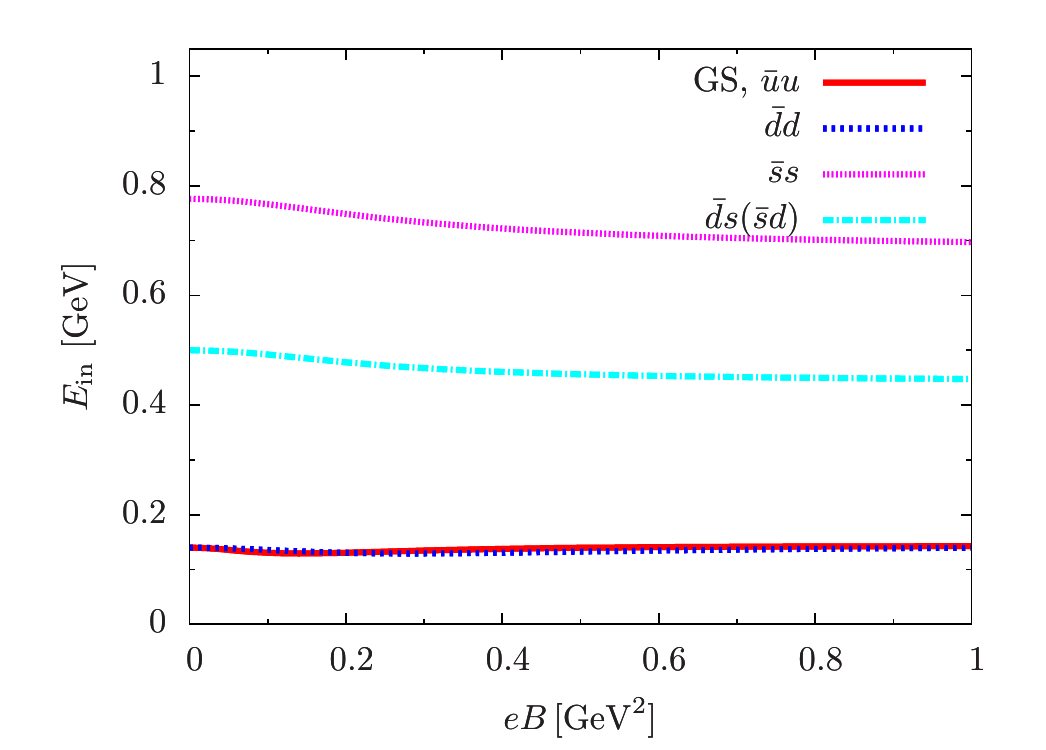}
}
\caption{The $B$-dependence of the ground state (GS) energy 
 $E_{\rm in}$ for charge neutral states with $(n_\perp ,l, n_z) =(0,0,0)$ and $ |S_z=0^- \ra $. 
Various flavor channels are considered. 
We neglected the $\bar{q}q$ annihilations.
}
\label{fig:neutral_GS_flavor_dep}       
\end{figure}

\begin{figure}
\resizebox{0.5\textwidth}{!}{%
  \includegraphics{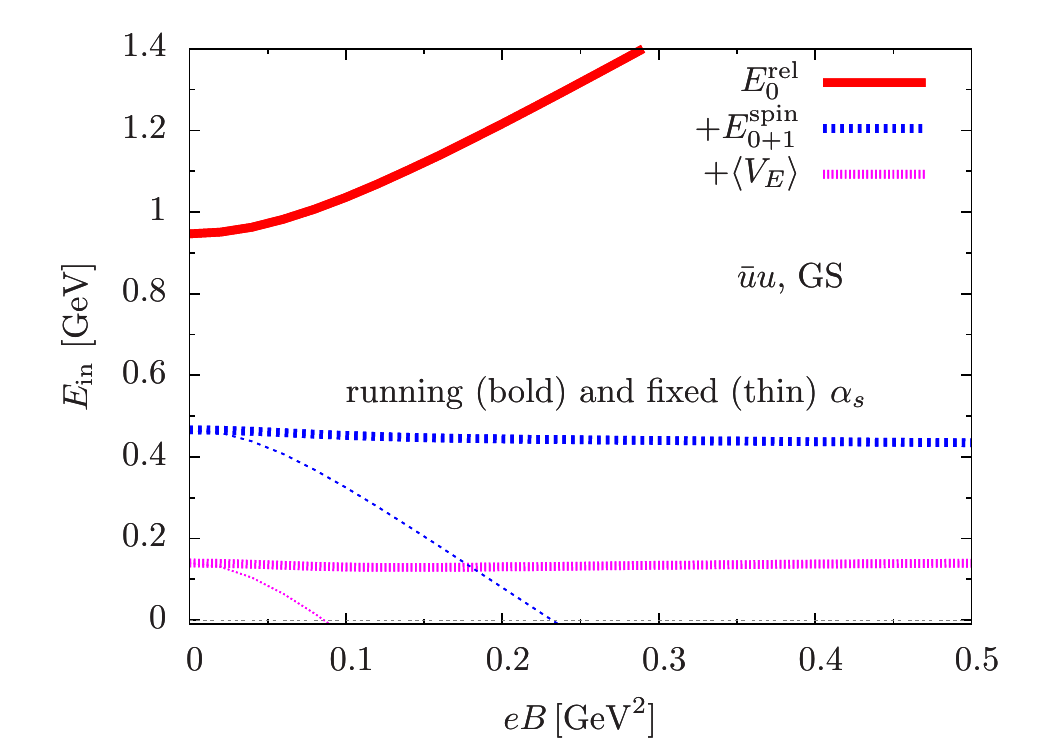}
}
\caption{ 
The $B$-dependence of the energy budget in the ground state (GS) for the $\bar{u}u$ channel.
 The $B$-dependent part of $E_{0}^{\rm rel}$ is largely cancelled by the Zeeman energy in $E_{0+1}^{\rm spin}$, and the attractive electric perturbation $\la V_E \ra$ reduces the energy.
We also plot the results of a fixed $\alpha_s$ for which mesons are unstable at large $B$. }
\label{fig:neutral_GS_energy_budget}       
\end{figure}

\begin{figure}
\resizebox{0.5\textwidth}{!}{%
  \includegraphics{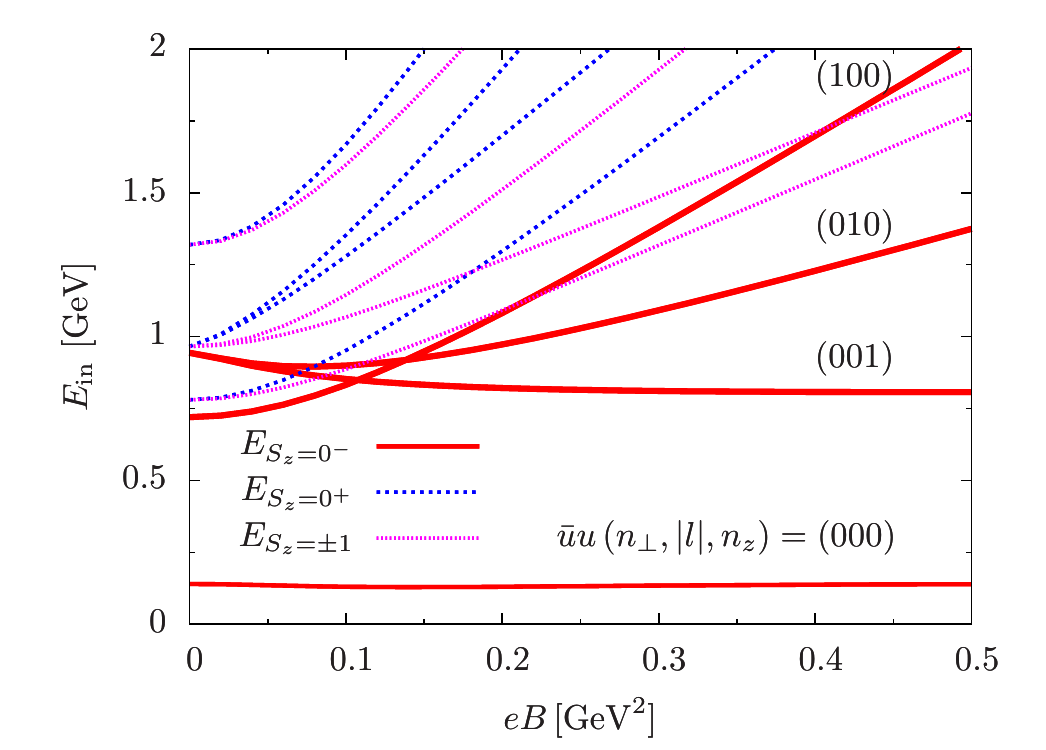}
}
\caption{ 
The $B$-dependence of the 1st excited states for the $\bar{u}u$ channel. One of $(n_\perp, |l|, n_z)$ is excited, and we attached $(n_\perp, |l|, n_z) = $(100), (010), (001) only to curves for the $S_z=0^-$ states. At $B=0$, the $S_z=0^-$ states become the spin singlet states, while the others form the spin triplet. }
\label{fig:neutral_1st_excited}       
\end{figure}

\subsection{Spectrum: numerical results}
\label{sec:numerical_neutral}

Now we examine the spectra of neutral mesons.
Shown in Fig.\ref{fig:neutral_GS_flavor_dep} are the ground state energies for $\bar{u}u$, $\bar{d}d$, $\bar{s}s$, $\bar{d}s$ ($\bar{s}d$) mesons at $\vK=0$. 
At finite $B$, the eigenstate is the mixture of the singlet and triplet states with $S_z=0$. 
Increasing $B$ leads to the reduction of the masses for small $B$, and to slightly increasing behaviors at very large $B$. 
The initial mass reduction is largely due to the cancellation of the zero point energy in the transverse dynamics and the Zeeman energy; 
at $B=0$ the zero point energy in confining potential is $\sim 2\times \sqrt{  2\alpha / \mu \,} $, 
while at large $B$ it becomes $\sim \calB_q/2\mu \sim B_q/2\mu + 4\mu \alpha /B_q$ whose leading term, $B_q/2\mu$, is cancelled by the Zeeman energy, leaving the small energy correction of $\sim \mu \alpha /B_q$. 

In Fig.\ref{fig:neutral_GS_energy_budget} we check the energy budget of $\bar{u}u$-mesons in the ground state. 
The other $B$-dependence comes from the modifications of wavefunctions which impact the evaluation of $\la V_{E,M} \ra$. 
We plot the results for a fixed $\alpha_s |_{B=0}$ with thin lines.
The matrix elements $\la V_{E,M} \ra$ are both negative.
As $B$ increases, for a running $\alpha_s$ its magnitude becomes insensitive to $B$, while for a fixed $\alpha_s$ the magnitude grows with $B$ and makes mesons unstable.

Next we examine the 1st excited states for the $\bar{u}u$ channel with various values of $S_z$. 
One of $(n_\perp, |l|, n_z)$ is excited. 
As illustrations, we attached the indices $(n_\perp, |l|, n_z) = $(100), (010), (001) to the curves for the $S_z = 0^-$ states. 
An excitation in these quanta not only costs the kinetic energy but also reduces the energy gain from the short range attractions. 
At large $B$, excitations in $n_\perp$ and $|l|$ cost large energies of $\sim |B|/\mu$, while excitations in $n_z$ are insensitive to $B$.

%

\section{Charged mesons}
\label{sec:charged}

The treatment of charged mesons is considerably different from the neutral meson case. Below we assume that the total charge is positive; this is realized only when both particles 1 and 2 have the positive charges (e.g. $u \bar{d}$ with $e_1=2/3$ and $e_2=1/3$). 
The nontrivial constants of motion are $\calK_\perp^2$ and $L_z$, both quantized. The dynamics is much more complicated than in the neutral meson case\footnote{A special simplification occurs when the conditions $e_1=e_2$ and $m_1 = m_2$ are both satisfied, as in quantum Hall systems made by many electrons. These conditions are not satisfied for mesons in QCD, but there may be some applications for diquarks with identical flavors.}.

\subsection{Unperturbed bases}
\label{sec:chargedl_unpert}

The first nontrivial step is the choice of coordinates. 
We use the center of mass coordinates for the $z$-direction as in the neutral meson case, but we apply different coordinates for the transverse directions;
we use the ``center of charge" coordinates ($e_R=e_1 +e_2$),
\beq
\vR_\perp = \frac{\, e_1 \vr_{1 \perp} + e_2 \vr_{2 \perp} \,}{e_R} \,,
\eeq
and the relative coordinate as $\vr = \vr_1 - \vr_2$.

Now we consider a set of operators $(\hat{\vPi}_R, \hat{\vPi}_r, \hat{\vcalK}_R, \hat{\vcalK}_r)$ such that $[ \hat{\vPi}_R, \hat{\vPi}_r]=0=[ \hat{\vcalK}_R, \hat{\vcalK}_r]$. 
First we look at a conserved operator $\hat{\vcalK}_R \equiv \hat{\vcalK}_1 + \hat{\vcalK}_2$,
\beq
\hat{\vcalK}_R
&=& \hat{\vp}_1 + \hat{\vp}_2 + \frac{\, \vB \,}{2} \times ( e_1 \hat{\vr}_1 + e_2 \hat{\vr}_2 ) 
\nonumber \\
&=& \hat{\vP}_R + \frac{\, 1 \,}{2} \vB_R \times \hat{\vR} \,.
\eeq
with $ \hat{\vP}_R = - \rmi \partial / \partial \vR $.
As for $ \hat{\calK}_r$, the condition $[ \hat{\calK}_R, \hat{\calK}_r]=0$ is satisfied for the choice 
(reduced charge: $e_r^{-1} = e_1^{-1} + e_2^{-1} = e_R/e_1 e_2$ and $\vB_r \equiv e_r \vB$)
\beq
\hat{\vcalK}_r 
\equiv \frac{\,e_2 \hat{\vcalK}_1 - e_1 \hat{\vcalK}_2 \,}{e_R}
= \hat{\vp}_r + \frac{\, 1 \,}{2} \vB_r \times \hat{\vr} \,,
\eeq
with $\hat{\vp}_r = - \rmi \partial/ \partial \vr $.
The $\hat{\vPi}_R$ and $ \hat{\vPi}_r$ are obtained by changing the signs, $\vB_R \rightarrow - \vB_R$ and $\vB_r \rightarrow - \vB_r$, in $\vcalK_R$ and $\vcalK_r$, respectively. 

Next, we consider an operator for a total (orbital) angular momentum,
\beq
\hat{\vL} = \sum_{j=1,2}  \hat{\vr}_j \times \hat{\vp}_j
= \hat{\vR} \times \hat{\vp}_R + \hat{\vr} \times \hat{\vp}_r = \hat{\vl}_R + \hat{\vl}_r \,,
\eeq
whose $z$-component is conserved; below we write $\hat{L}_z = ( \hat{l}_R + \hat{l}_r )_z $. 
The eigenvalues of $(\hat{\vcalK}_R^2, \hat{\vPi}_R^2, \hat{l}_R)$ are labeled with indices $(N_\calK, N_\Pi, l_R)$.

Now we rewrite our hamiltonian in our new coordinates. Treating the dynamics in the $z$-direction as before, the unperturbed hamiltonian is
\beq
\hat{H}_0
=  \frac{\, K_z^2 \,}{\, 2M \,}
+ M 
+ E_0^{n_z}
+ \hat{H}_0^\perp
- \hat{ \vec{\mu} } \cdot \vB
\,,
\eeq
where the transverse part is
\beq
\hat{H}_0^\perp =  
c_R \hat{\vPi}_R^2
+ c_{\rm mix} \hat{\vPi}_R \cdot \hat{\vPi}_r 
+ \frac{\, \hat{\vPi}_r^2 \,}{\, 2 \mu \,} 
+ \alpha \hat{ \vr }_\perp^2  \,,
\eeq
with the coefficients
\beq
c_R = \frac{1}{\, 2 e_R^2 \,} \bigg( \frac{ e_1^2 }{\, m_1 \,} +  \frac{ e_2^2 }{\, m_2 \,}  \bigg) \,,~~
c_{\rm mix} = \frac{1}{\,  e_R \,} \bigg( \frac{ e_1 }{\, m_1 \,} -  \frac{ e_2 }{\, m_2 \,}  \bigg) \,.
\nonumber \\
\eeq
The analyses of $\hat{H}_\perp$ become complicated\footnote{
The exception is the case of identical particles, $e_1=e_2=e_R/2$ and $m_1=m_2=M/2$, for which $c_{\rm mix}=0$ and $c_R = 1/2M$, which allow us to separately treat $R$- and $r$-parts.
} due to the coupling term $\hat{\vPi}_R \cdot \hat{\vPi}_r$. 

We work with the bases, 
$\big| N_\Pi; n_{\tilde{\Pi} }, n_{\tilde{\calK} } \big\ra = |N_\Pi  \ra \otimes | n_{\tilde{ \Pi} }, n_{ \tilde{\calK} } \ra$, 
which are the eigenstates of $\hat{\vPi}_R^2$ and $\hat{\vPi}_r^2/2\mu + \alpha \hat{\vr}_\perp^2$.
These operators can be expressed by the creation and annihilation operators, see Sec.\ref{sec:rearrangement}. 
With some algebras, we readily find
\beq
\hat{\Pi}_R^2 |N_\Pi  \ra = |B_R| ( 2 N_\Pi + 1 ) |N_\Pi \ra \,.
\eeq
Meanwhile, the term with the subscript $r$ requires some effort to find the eigenstates. First we rewrite
\beq
\hat{\vPi}_r^2 + 2\mu \alpha \hat{\vr}_\perp^2 
= \hat{ \tilde{\vPi} }_r^2 + ( \vec{\calB}_r - \vB_r ) \cdot \hat{\vl }_r\,,
\label{eq:tPi_r^2}
\eeq
where $\calB_r^2 \equiv B_r^2 + 8 \mu \alpha$ with the direction $\vec{ \calB}_r \parallel \vB_r$, and
\beq
\hat{ \tilde{\vPi} }_r 
\equiv \hat{ \vp }_r  - \frac{1}{\, 2 \,} \vec{\calB}_r \times \hat{ \vr } 
 \,,
\eeq
which is obtained by replacement, $\vec{B}_r\rightarrow \vec{\calB}_r$ in $\hat{ \vec{\Pi} }_r$. 
Also $\hat{ \tilde{ \vec{\calK} } }_r$ is obtained from $\hat{ \vec{\calK} }_r$ in the same way. 
As $\hat{l}_z \propto   \hat{ \tilde{ \vec{\calK} } }_r^2 - \hat{  \tilde{ \vec{ \Pi} } }_r^2$,
Eq.(\ref{eq:tPi_r^2}) can be diagonalized as
\beq
&& \bigg( \frac{\,  \hat{ \vPi }_r^2 \,}{\, 2 \mu \,} + \alpha \hat{\vr}_\perp^2 \bigg)
| n_{\tilde{ \Pi} }, n_{ \tilde{\calK} } \ra
 \nonumber \\
 && =  \frac{\, 1 \,}{\, 2 \mu \,} \big[\, |\calB_r| \big( 2 n_{ \tilde{\Pi} } + 1 \big) 
 	+ (\calB_r - B_r ) l_r \, \big] \, | n_{\tilde{ \Pi} }, n_{ \tilde{\calK} } \ra \,,
\eeq
where $l_r$ can be expressed as 
$l_r =  n_{\tilde{\calK}} - n_{\tilde{\Pi} } $, see Eq.(\ref{eq:l_and_nPi}).

We also need to evaluate the cross terms or off-diagonal elements.
We rewrite the expression in terms of $( \hat{\Pi}_R^{\pm}$, $\hat{\tilde{\Pi}}^\pm_r, \hat{\tilde{\calK}}^\pm_r$), 
where $\hat{\Pi}^\pm = \hat{\Pi}^x \pm \rmi \hat{\Pi}^y $, 
$\hat{ \calK }^\pm = \hat{ \calK }^x \pm \rmi \hat{\calK }^y $, etc.,
\beq
&& \hat{\vPi}_R \cdot \hat{\vPi}_r 
= \frac{1}{\, 2 \,} \big( \hat{\Pi}_R^+ \hat{\Pi}_r^- + \hat{\Pi}_R^- \hat{\Pi}_r^+ \big) 
\nonumber \\
&&=  \frac{\, f_+ \,}{\, 2 \,} \big( \hat{\Pi}_R^+ \hat{ \tilde{\Pi} }^-_r + \hat{\Pi}_R^- \hat{ \tilde{\Pi} }^+_r \big)
 +   \frac{\, f_- \,}{\, 2 \,} \big( \hat{\Pi}_R^+ \hat{ \tilde{\calK} }^-_r + \hat{ \Pi}_R^- \hat{ \tilde{\calK} }^+_r \big)
\,,
\nonumber \\
\label{eq:pi_R_pi_r}
\eeq
with $f_\pm \equiv 1\pm B_r/\calB_r $,  see Sec.\ref{sec:rearrangement} for more details.
The overall scale is $\hat{ \vPi}_R \cdot \hat{\vPi}_r \sim \sqrt{ | B_R \calB_r|}$. 

The off-diagonal elements are calculated from the 
relation\footnote{Some qualitative features. 
For the weak $B$ case, $B_r /\calB_r \sim B_r /\lqcd^2 \ll 1$ so that the coupling behaves as $f_\pm/2 \sim 1/2$. 
The first two terms include $\tilde{\Pi}_r^\pm$ which, at weak $B$, mainly describe the excitations inside of confining potentials, 
while the last two terms with $\tilde{\calK}_r^\pm$ describe the motion of the guiding centers in relative coordinates. 
For the strong $B$ case,  $B_r /\calB_r \sim B_r /\lqcd^2 \simeq 1-\lqcd^4/B_r^2$, so $f_- \sim \lqcd^4/B_r^2 \ll 1$. 
In this regime excitations within the confining potential can easily occur with $f_+ \sim 1$, while the processes involving changes in $n_{\tilde{\calK}}$ are suppressed by a factor $f_-$.
}
\beq
&&\frac{\, \hat{\vPi}_R \cdot \hat{\vPi}_r \,}{\, \sqrt{ |B_R| |\calB_r | } \,} \, \big| N_\Pi ; n_{\tilde{\Pi} }, n_{\tilde{\calK} } \big\ra
\nonumber \\
&&
=  f_+ \sqrt{ (N_\Pi +1) n_{\tilde{\Pi} } \,} \, \big| N_\Pi +1 ; n_{\tilde{\Pi} } -1, n_{\tilde{\calK} } \big\ra
\nonumber \\
&&
~+ f_+  \sqrt{ N_\Pi ( n_{\tilde{\Pi} } +1 ) \,} \, \big| N_\Pi -1 ; n_{\tilde{\Pi} } +1, n_{\tilde{\calK} } \big\ra
\nonumber \\
&&
~ +  f_-  \sqrt{ (N_\Pi +1)( n_{\tilde{\calK} } +1 ) \,} \, \big| N_\Pi + 1 ; n_{\tilde{\Pi} } , n_{\tilde{\calK} } + 1\big\ra 
\nonumber \\
&&
~ +  f_-  \sqrt{ N_\Pi n_{\tilde{\calK} } \,} \, \big| N_\Pi - 1 ; n_{\tilde{\Pi} } , n_{\tilde{\calK} } - 1\big\ra 
\,.
\label{eq:offdiagonal}
\eeq
Here we summarize the relations among indices. 
We recall that the conserved numbers are $N_{\calK}$ and $L_z = l_r + l_R$
where $l_r = n_{\tilde{\calK}} - n_{\tilde{\Pi}}$ and $l_R = N_\calK - N_\Pi$.
We assume $N_\calK$ and $L_z$ are given. 
Then we can take $N_\Pi$ and $n_{\tilde{\Pi}}$ as independent variables, while $n_{\tilde{\calK}}$ can be expressed by the variables ($N_\Pi$,$n_{\tilde{\Pi}}$) and the conserved numbers $(N_\calK, L_z)$.
Actually, it turns out that the spectra depends on $N_\calK$ and $L_z$ only through the combination
\beq
\calL_z \equiv L_z - N_\calK
~~~\big(= n_{\tilde{\calK} } - n_{\tilde{\Pi} } - N_{\Pi} \big) \,.
\eeq
Indeed, the quantum numbers, 
$n_{\tilde{\calK} } = n_{\tilde{ \Pi} } + l_r $ and $l_r$, 
can be expressed by $(N_\Pi, n_{\tilde{\Pi}}; \calL_z)$,
\beq
l_r &=& L_z - l_R = L_z - (N_{\calK} - N_\Pi )
= \calL_z + N_{\Pi}  \,.
\label{eq:quanta_charged}
\eeq
We write the energy eigenstate by the linear combination of the bases 
\beq
&& \big| N_\Pi , n_{\tilde{\Pi} } \ra\ra_{ \calL_z } 
\nonumber \\
&& \equiv \big| N_\Pi \ra \otimes | n_{\tilde{\Pi} }, n_{\tilde{\calK} } =n_{\tilde{\Pi} } + N_\Pi + \calL_z \big\ra.
\eeq
With this basis, we can simplify Eq.(\ref{eq:offdiagonal}) as
\beq
&&\frac{\, \hat{\vPi}_R \cdot \hat{\vPi}_r \,}{\, \sqrt{ |B_R| |\calB_r | } \,} \, \big| N_\Pi , n_{\tilde{\Pi} } \ra\ra_{ \calL_z } 
\nonumber \\
&&
=  f_+ \sqrt{ (N_\Pi +1) n_{\tilde{\Pi} } \,} \, \big| N_\Pi +1 , n_{\tilde{\Pi} } -1 \ra\ra_{ \calL_z } 
\nonumber \\
&&
~+ f_+  \sqrt{ N_\Pi ( n_{\tilde{\Pi} } +1 ) \,} \, \big| N_\Pi -1, n_{\tilde{\Pi} } +1  \ra\ra_{ \calL_z } 
\nonumber \\
&&
~ +  f_-  \sqrt{ (N_\Pi +1)( n_{\tilde{\calK} } +1 ) \,} \,\big| N_\Pi +1, n_{\tilde{\Pi} } \ra\ra_{ \calL_z } 
\nonumber \\
&&
~ +  f_-  \sqrt{ N_\Pi n_{\tilde{\calK} } \,} \, \big| N_\Pi - 1, n_{\tilde{\Pi} } \ra\ra_{ \calL_z } 
\,.
\label{eq:offdiagonal_revised}
\eeq
with $n_{\tilde{\calK} } =n_{\tilde{\Pi} } + N_\Pi + \calL_z$.
Thus, the dimensions of the vectors are determined from the product of the dimensions for $N_\Pi$ and $n_{\tilde{\Pi}}$.

In summary, we label the eigenstates of $\hat{H}_0^\perp$ as
\beq
| E_0^\perp ; \calL_z \ra 
= \sum_{N_\Pi, n_{\tilde{\Pi} } =0 } C^{ N_\Pi,\, n_{\tilde{\Pi} } }_{ E_0^{\perp}  } \big| N_\Pi , n_{\tilde{\Pi} } \ra\ra_{ \calL_z }  \,,
\label{eq:E0_perp_c_N}
\eeq
for a given $\calL_z$ (and $ N_\calK$ if we wish to write everything explicitly). As we have just mentioned, the $N_\calK$ does not affect the spectrum.
So we omit $N_\calK$ from the label.
The coefficients $C^{ N_\Pi,\, n_{\tilde{\Pi} } }_{ E_0^\perp }$ are determined by numerical diagonalization.
In practice, it is useful to note that the condition $N_\calK, N_\Pi, n_{ \tilde{\calK} }, n_{ \tilde{\Pi} } \ge 0$ put a constraint,
\beq
n_{ \tilde{\calK} } \ge 0 
~\rightarrow~  \calL_z \ge - \big( n_{\tilde{ \Pi} } + N_{\Pi} \big) \,.
\label{eq:NcalKL_upperbound}
\eeq
Thus if $\calL_z < 0$, then at least either $n_{\tilde{ \Pi} }$ or $N_{\Pi}$ must have the excitation, costing energy of $\sim \sqrt{ |B_r| }$ or $\sim \sqrt{ |B_R| }$.

To summarize, our unperturbed hamiltonian for the bases $ | n_z \ra \otimes | E_0^{\perp}, \calL_z \ra $ is
\beq
\hat{H}_0
=  E_K + E_0^{\rm rel} - \hat{ \vec{\mu} } \cdot \vB
\,,
\eeq
with
\beq
 E_K =  \frac{\, K_z^2 \,}{\, 2M \,}  \,, ~~~~~
 E_0^{\rm rel} &=&  M + E_0^{n_z} + E_0^{\perp,\, \calL_z}  \,, 
\eeq
where $ \hat{H}_0^\perp | E_0^{\perp}, \calL_z \ra =  E_0^{\perp,\, \calL_z} | E_0^{\perp}, \calL_z \ra$.

\subsection{Perturbations}
\label{sec:charged_pert}

As in the neutral meson case, we treat short range correlations as perturbations (see Sec.\ref{sec:short}). 
For $Q_\perp^2$ in $\alpha_s(Q^2)$ (see Eq.(\ref{eq:Q2})), we use $Q_\perp^2 = |\calB_r|$.
Using the unperturbed bases $|n_z;K_z\ra \otimes |E_0^\perp; \calL_z \ra$, 
we estimate the hamiltonian as
(see Eq.(\ref{eq:E_K}))
 \beq
 \la \big( \hat{H}_0 + \hat{H}_1 \big) \ra
 = E_K + \hat{H}_{\rm in} \,,
 \label{eq:E_in_ch}
 \eeq
 where $\hat{H}_{\rm in} \equiv E_{0+1}^{\rm rel} +  \hat{H}_{0+1}^{\rm spin} $ with 
\beq
E_{0+1}^{\rm rel}
&=& E_{0}^{\rm rel} + \la V_E \ra \,,  
	\nonumber \\
\hat{H}_{0+1}^{\rm spin}
&=&  - \hat{ \vec{\mu} }  \! \cdot \!\vB + \la V_s \ra \,  \hat{ \vec{\sigma} }_1 \cdot \hat{ \vec{\sigma} }_2  \,.
\label{eq:H_spin_ch}
\eeq
In the next section we find the eigenstates for $\hat{H}_{0+1}^{\rm spin}$.

The orbital matrix element is (see Eq.(\ref{eq:E0_perp_c_N})), 
\beq
\la V \ra^{\calL_z}_{E_0^\perp, n_z}
&&= \sum_{N_\Pi, n_{\tilde{\Pi} }, n'_{\tilde{\Pi} }  } C^{ N_\Pi,\, n_{\tilde{\Pi} } }_{E_0^{\perp}  } C^{ N_\Pi,\, n'_{\tilde{\Pi} } }_{E_0^{\perp}  }
\nonumber \\
&&\times \, \la n_z, N_\Pi , n_{\tilde{\Pi} } ;  \calL_z \big| V \big| n_z, N_\Pi , n'_{\tilde{\Pi} } ;  \calL_z \big\ra 
 \,,
\eeq
where we have used the fact that $N_\Pi$ is diagonal for $V$ depending only on the relative coordinate $\hat{\vr}$.
In practice, the matrix element is evaluated as
\beq
&& \hspace{-0.1cm} 
\la n_z, N_\Pi , n_{\tilde{\Pi} } ;  \calL_z \big| V \big| n_z, N_\Pi , n'_{\tilde{\Pi} } ;  \calL_z \big\ra 
 \nonumber \\
&&
= \int \rmd z \rmd^2 \vr_\perp \, | \psi_{n_z} (z)|^2 \, (\Phi_{\calB_r}^*)_{n_\perp}^l \Phi_{\calB_r, n'_\perp}^{l'} (\vr_\perp) V(\vr)  \,,
\label{eq:exp_value_charged}
\eeq
where the definitions of $\psi_{n_z}$ and $\Phi_{B, n_\perp}^{l} $ were given in Secs.\ref{sec:formulae_z}, and
\beq
l = n_{\tilde{\calK}} - n_{\tilde{\Pi}}  \,,~~~~ 
n_\perp = \left\{ \begin{array}{l}
~ n_{\tilde{\Pi}} ~~~~~~\!(l \ge 0) \\
~ n_{\tilde{\calK}} ~~~~~~(l < 0) 
\end{array} \right. 
\eeq
with $n_{\tilde{\calK} } =n_{\tilde{\Pi} } +N_\Pi + \calL_z$.
The $l'$ and $n_\perp'$ can be obtained by replacement $n_{\tilde{\Pi}} \rightarrow n'_{\tilde{\Pi}}$.

\subsection{Spin dependent terms}
\label{sec:quanta}

The Zeeman splitting terms are given by
\beq
- \hat{ \vec{\mu} } \cdot \vB
=   \frac{\, B \,}{2}  
\times
\left\{ \begin{array}{l}
- \big(\frac{e_1}{m_1} + \frac{e_2}{m_2}\big) ~~~~(\up \up) \\
+ \big(\frac{e_1}{m_1} + \frac{e_2}{m_2}\big) ~~~~(\down \down) \\
+ \big(\frac{e_1}{m_1} - \frac{e_2}{m_2}\big) ~~~~(\down \up) \\
- \big(\frac{e_1}{m_1} - \frac{e_2}{m_2}\big) ~~~~(\up \down) 
\end{array} \right.
\eeq
At very large positive (negative) $e_{1} B$ and $e_2 B$, the $|\! \up \up\ra$ ($|\!\down \down\ra$) state tends to cancel the zero point energy from $\hat{\vPi}_R^2$ and $\hat{\vPi}_r^2$ terms,
\beq
 \bigg( c_R \hat{ \vPi }_R^2 + \frac{\, \hat{ \vPi }_r^2 \,}{\, 2\mu \,} \bigg)_{N_\Pi = n_{\tilde{\Pi} } = 0 }
 = \frac{\, B \,}{2} \bigg( \frac{ e_1 }{\, m_1 \,} +  \frac{ e_2 }{\, m_2 \,}  \bigg)\,,
\eeq
and hence the ground state energy at large $B$ becomes insensitive to $B$. 
As a result, the charged $\rho$-mesons, $\rho_+^{S_Z=1}$ and $\rho_-^{S_Z=-1}$, become the ground states,
while the energies of the other states such as $\pi_\pm$ and $\rho_+^{S_z=-1,0}$ are lifted up.

For a general $B$, as before the spin dependent part is treated within the first order perturbation theory,
%
\beq
E_{0+1}^{\rm spin} 
= \la V_s (r) \ra \la \hat{ \vec{\sigma} }_1 \cdot \hat{ \vec{\sigma} }_2 \ra_{\rm spin} - \la \hat{ \vec{\mu} } \ra_{\rm spin} \cdot \vB  \,.
\eeq
where $\la V_s \ra$ is the expectation value for the state $| n_{E_\perp}; \calL_z \ra $ determined from the zeroth order hamiltonian.

The spin-aligned components are diagonal,
%
\beq
\big(E^{\rm spin}_{0+1} \big)_{S=1, S_z= \pm 1}
= \la V_s \ra \mp \frac{\, B \,}{\, 2 \,}  \bigg(\frac{e_1}{m_1} + \frac{e_2}{m_2} \bigg) \,,
\eeq
while, for the $S_z=0$ components, we diagonalize the matrix
\beq
\left[
\begin{matrix}
~~  \la V_s \ra &~~  -\frac{\, B \,}{\, 2 \,}  \big(\frac{e_1}{m_1} - \frac{e_2}{m_2} \big)  ~~  \\
~~  - \frac{\, B \,}{\, 2 \,}  \big( \frac{e_1}{m_1} - \frac{e_2}{m_2} \big)  ~&~ -3 \la V_s \ra ~~\\
\end{matrix} \right]
\nonumber 
\eeq
which leads to the eigenvalues
\beq
&&\big(E^{\rm spin}_{0+1} \big)_{S_z= 0^{\pm } }
\nonumber \\
&& = - \la V_s \ra  \pm \sqrt{ 4 \la V_s \ra^2 + \frac{\, B^2 \,}{\, 4 \,}  \bigg( \frac{e_1}{m_1} - \frac{e_2}{m_2} \bigg)^2 \,} \,.
\eeq
%

\subsection{Orbital excitations; trends in the decoupling limit}
\label{sec:some_examples}

To get qualitative insights we discuss some examples for low energy states. 
We consider the decoupling limit where the off-diagonal terms in Eq.(\ref{eq:offdiagonal}) are neglected.
Then, the hamiltonain depends on the quanta $N_\Pi$, $n_{\tilde{\Pi}}$, and $l_r$, 
which contribute to the energy terms, $\sim N_\Pi c_R |B_R|$, $ \sim n_{\tilde{\Pi}} |\calB_r|/2\mu$, 
and $ (\calB_r - B_r) l_r/2\mu$, respectively. 

In the weak field regime, $|B_R| ( \ll \lqcd^2)$ is negligible and the details of $N_\Pi$ are not important. 
Meanwhile $|\calB_r| \sim \lqcd^2$ dominates the dynamics and nonzero $n_{\tilde{\Pi}}$ or $l_r$ cost the energy of $\sim \lqcd$. 
Therefore the low-lying states consist of $n_{\tilde{\Pi}}=l_r =0$ (see Eq.(\ref{eq:quanta_charged})), for which 
$\calL_z = - N_\Pi \le 0$,
and $N_\Pi$ induces small energy splittings of $\sim |B_R|$ within the $n_{\tilde{\Pi}}=l_r =0$ states. 
For $B\rightarrow 0$, their spectra together form the center of mass energy of the form $\sim \vK_\perp^2/2\mu$ in the transverse directions. 

In the strong field regime, $|B_R| $ and $|\calB_r|$ are large $\sim |B| (\gg \lqcd^2)$, so that the low energy states must have $N_\Pi=n_{\tilde{\Pi}} =0$. 
In this case Eqs.(\ref{eq:quanta_charged}) and (\ref{eq:NcalKL_upperbound}) requires 
$l_r = \calL_z \ge 0$, which costs the energy $ (\calB_r - B_r) l_r/2\mu$ (for $B>0$), but at large $B$ it is small,
$ (\calB_r - B_r) l_r/2\mu \sim  \lqcd^4/|B| \times l_r/\mu$.
Hence the spectra is insensitive to $l_r$ at large $B$ (except for a very large $l_r$). 

Including the coupling $\hat{ \vPi}_R \cdot \hat{\vPi}_r $ makes the analyses more complicated,
but the discussions above appear to give the good baselines (Fig.\ref{fig:charged_excited_calL}).

\subsection{Numerical results}
\label{sec:numerical_charged}

\begin{figure}[thb]
\resizebox{0.5\textwidth}{!}{%
  \includegraphics{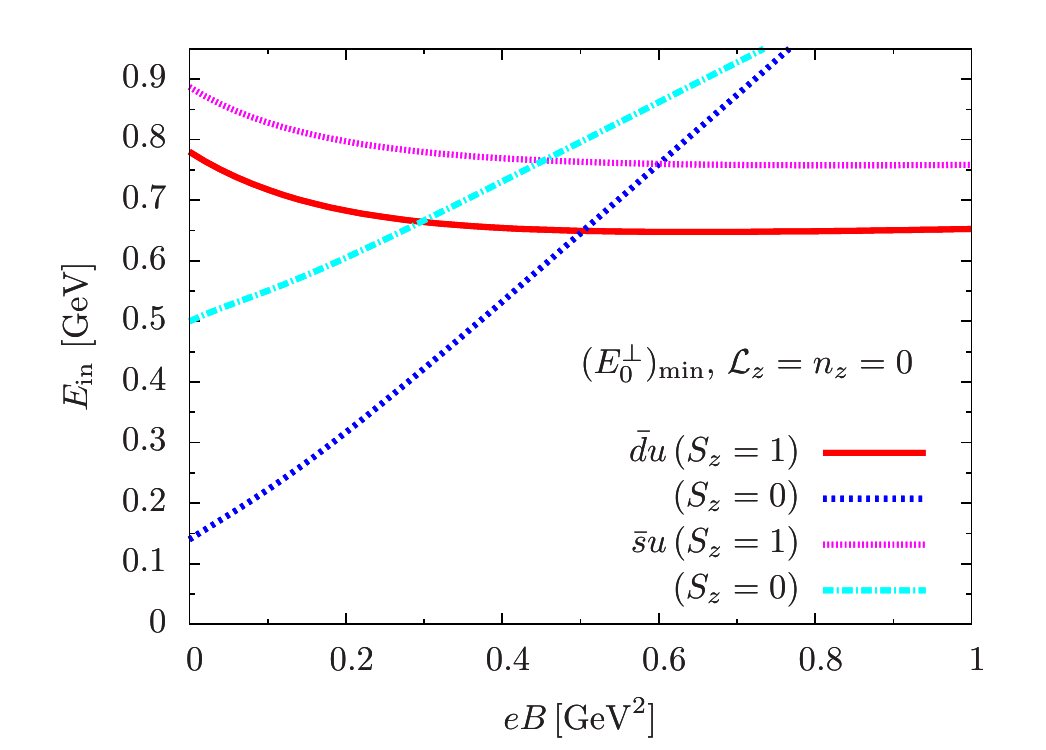}
}
\caption{ 
The $B$-dependence of the energy $E_{\rm in}$ for the $ (E_0^{\perp } )_{\rm min}$ and $ \calL_z = n_z = 0$ states in the $\bar{d}u$ ($\bar{s}u$) channel with $S_z=$ 0 and 1, which at $B=0$ correspond to $\pi_+$ ($K_+$) and $\rho_+$ ($K^*_+$) mesons. The mass of $\bar{d} u$ ($\bar{s}u$) with $S_z=-1$ degenerates with that of $\bar{u}d$ ($\bar{u}s$) mesons with $S_z=-1$. }
\label{fig:charged_GS}       
\end{figure}

\begin{figure}
\resizebox{0.5\textwidth}{!}{%
  \includegraphics{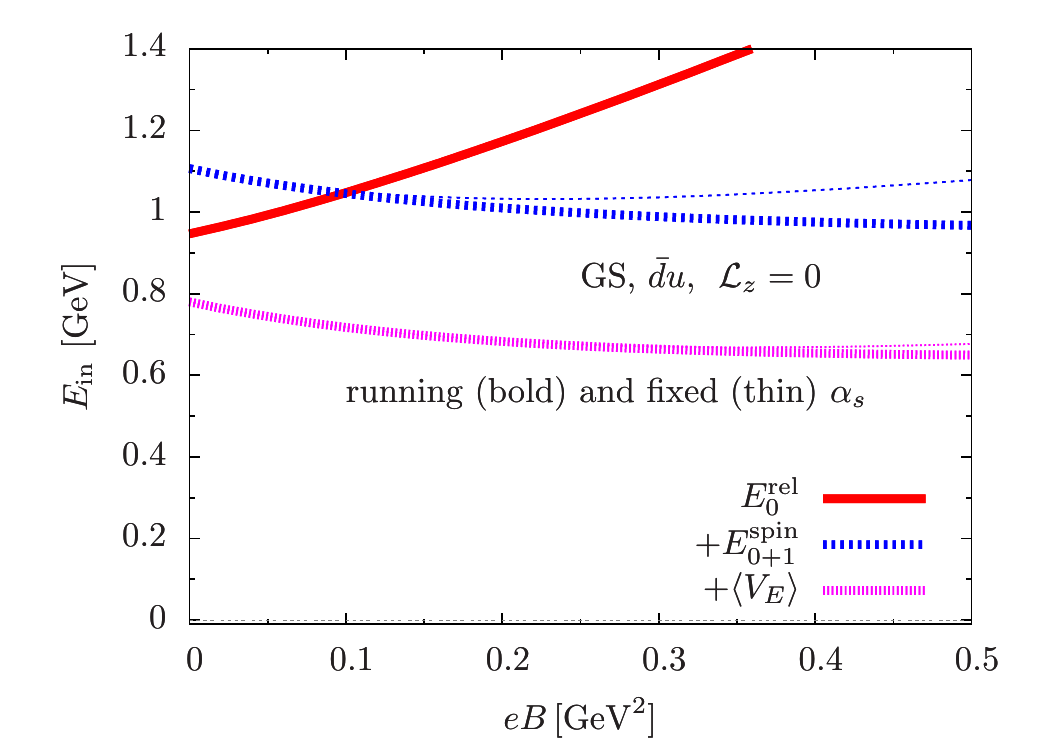}
}
\caption{ 
The $B$-dependence of the energy budget in the ground state (GS) with $\calL_z=0$. The $\bar{d}u$ channel is shown.
For this channel, the short range correlations $\la V_E \ra$ and $\la V_M \ra$ tend to cancel as they have the opposite sign.
 We also plot the results of a fixed $\alpha_s$. 
}
\label{fig:charged_Lz0_energy_budget}       
\end{figure}

\begin{figure}[thb]
\resizebox{0.5\textwidth}{!}{%
  \includegraphics{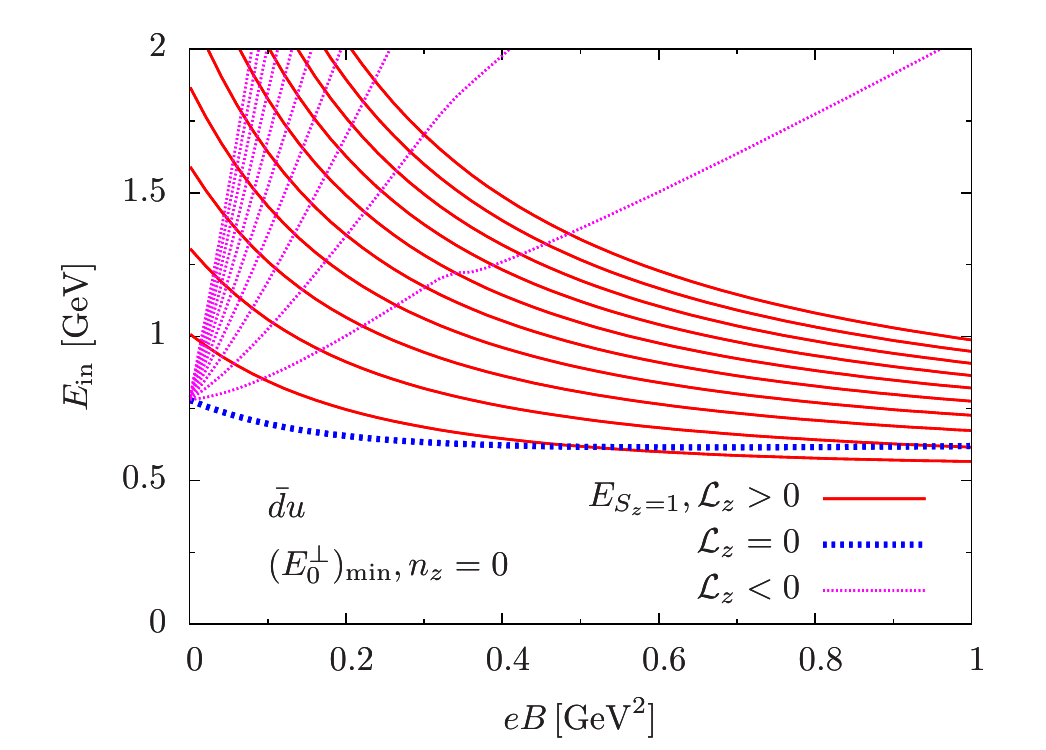}
}
\caption{ 
The $B$- and $\calL_z$-dependence of the energy $E_{\rm in}$ for the $ (E_0^{\perp } )_{\rm min}$  and $n_z = 0$ states in the $\bar{d}u$ ($\bar{s}u$) channel with $S_z= 1$. 
The quanta $\calL_z$ are displayed from $-10$ to $+10$. At small $B$, the low-lying states are dominated by the $l_r=0$ state and series of 
$\calL_z = - N_\Pi \le 0$ 
states are largely degenerated. At large $B$, magnetic fields favor the $n_{\tilde{\Pi} }= N_\Pi =0$ states as the low energy states, with which many spectra with  $l_r = \calL_z \ge 0$ appear at low energy.
}
\label{fig:charged_excited_calL}       
\end{figure}

First we examine the low energy states of charged mesons (Fig.\ref{fig:charged_GS}). Here we display only positively charged mesons, $\bar{d}u$ and $\bar{s}u$. 
(The results for $\bar{u}d$ and $\bar{u}s$ are obtained by flipping charges and spins at the same time.) 
At $B=0$, $\pi_+$, $K_+$, $\rho_+$, and $K_+^*$ are ground states for given quantum numbers. 
As $B\neq 0$, $(E_0^{\perp}, \calL_z, n_z)$ become good quantum numbers, and we examine the $(E_0^{\perp}, \calL_z, n_z)=( (E_0^{\perp})_{\rm min},0,0)$ case here. 
The energies of the $\pi_+$ and $K_+$ quantum numbers at $B=0$ are lifted up by magnetic fields. 
Meanwhile, the $\rho_+^{S_z=1}$ and $(K^*_+)^{S_z=1}$ states at $B=0$ have the energy reduction and their masses approach constant values at very large $B$. 
At some point the $S_z=1$ states become the ground states for the charged meson.

The energy budget in $\bar{d}u$ mesons is shown in Fig.\ref{fig:charged_Lz0_energy_budget}. 
Here  the ground state (GS) for $\calL_z=0$ is considered.
As in neutral mesons, for charged mesons the zero point and Zeeman energies tend to cancel.
Meanwhile, in contrast to the neutral meson cases, the short range correlations $\la V_E \ra$ and $\la V_M \ra$ have the opposite signs and hence tend to cancel.
We also show the results for a fixed $\alpha_s$ with the thin lines.
Unlike the neutral meson cases, the use of a fixed $\alpha_s$ does not lead to unstable modes for the range of $B$ we have explored;
both $\la V_E \ra$ and $\la V_M\ra$ grow in the magnitude but they largely cancel.
For a running $\alpha_s$ $\la V_E \ra$ and $\la V_M\ra$ do not change much for increasing $B$.

The behaviors of the excited states are considerably different at $B=0$ and $B \neq 0$. 
Shown in Fig.\ref{fig:charged_excited_calL} are the spectra of the $(E_0^{\perp})_{\rm min}$ and $n_z=0$ states for various $\calL_z$. 
As discussed in Sec.\ref{sec:some_examples},
at small $B$, excitations with 
$\calL_z \simeq - N_\Pi < 0$
cost small energies of $\sim N_\Pi |B|/2\mu$, and the series of $\calL_z$ form very dense spectra. 
Increasing $B$ turns them into discrete levels. 
On the other hand, excitations with 
$\calL_z \simeq l_r > 0$ form discrete spectra of $\sim l_r \lqcd$ at small $B$, 
and the energy splittings become closed at large $B$ as $\sim \lqcd^3/|B|$. 
Some states with $\calL_z > 0$ become less energetic than the $\calL_z=0$ state.

\section{Hadron resonance gas }
\label{sec:HRG}

\begin{figure}
\resizebox{0.5\textwidth}{!}{%
  \includegraphics{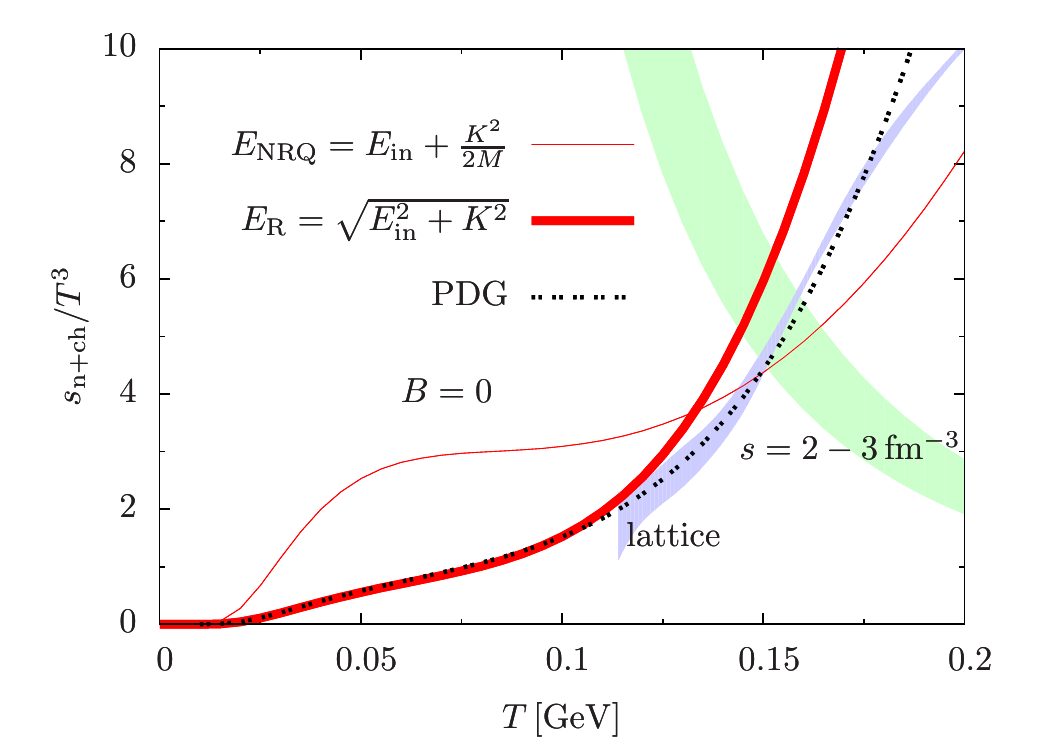}
}
\caption{
Normalized entropy density of the HRG model. 
Only mesons are included for our quark model.
The center of mass energy is treated in two ways;
the non-relativistic quark model (NRQ, thin line, red), and the phenomenological relativistic extension (R, bold line, red).
We also show the lattice result (including mesons {\it and} baryons) \cite{Bali:2014kia}, a band for constant entropy density $s=2-3\, {\rm fm}^{-3}$,
and the HRG result based on meson and baryon spectra from the Particle Data Group (PDG) \cite{ParticleDataGroup:2020ssz}.
}
\label{fig:total_s_Bdep_B000}       
\end{figure}

\begin{figure}
\begin{center}
\resizebox{0.4\textwidth}{!}{%
  \includegraphics{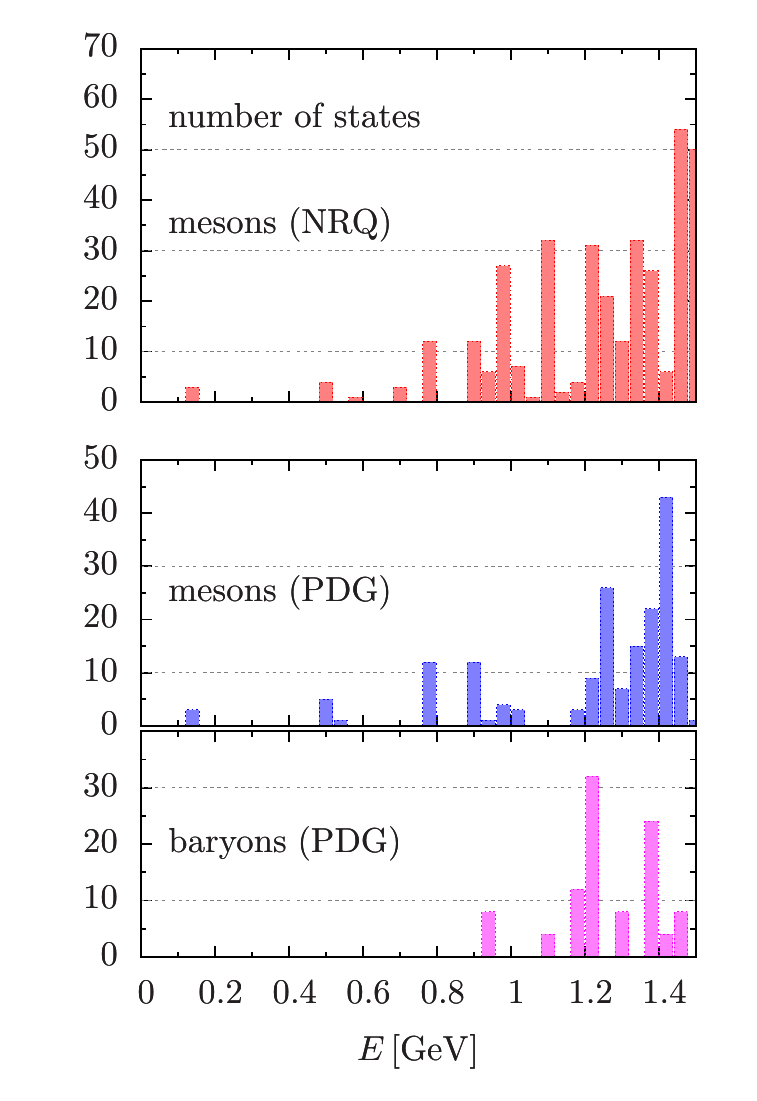}
}
\end{center}
\caption{
The number of hadronic states. 
The uppermost is for mesonic spectra in our quark model; the middle and lowermost for mesonic and baryonic spectra in the PDG, respectively.
Our quark model predicts too many mesonic states at $E \gtrsim 1$ GeV.
}
\label{fig:B0_histgram}       
\end{figure}

As in the usual HRG model \cite{Karsch:2003vd}, we apply the ideal gas description for mesons and calculate
the thermodynamic quantities as the sum of each mesonic contribution.
The description should be valid in dilute or low temperature regimes. 
It is known that the HRG model {\it with experimental hadron spectra} reproduces the lattice data quite well, 
up to the critical temperature, $T_c\simeq 156.5\pm 1.5$ MeV \cite{HotQCD:2019xnw}, where hadrons begin to overlap.
At finite $B$, such spectra are not available experimentally.
For this reason we use the hadron spectra computed in our quark model and then construct the HRG.
The results will be compared with the lattice results in Ref.\cite{Bali:2014kia}.

We compute only the thermal part of the pressure from neutral mesons as
\beq
P^{\rm th} (T, B) \equiv P (T, B) - P (T=0, B) \,,
\eeq
and will not directly address the issues related to the zero temperature part $P (T=0, B)$.
The latter requires dynamical determination of the effective quark masses which
are inputs rather than outputs in our non-relativistic quark models.

We note that our HRG does not contain baryons.
Therefore our HRG {\it must} underestimate the entropy density.
In the following results, we include the resonances whose rest masses are less than $2.5$ GeV. 
We have checked that resonances with higher energies do not affect the entropy density significantly at $T\lesssim 250$ MeV.

\subsection{The $B=0$ case }
\label{sec:HRG_B0}

We first examine how our predictions work at $B=0$.
Shown in Fig.\ref{fig:total_s_Bdep_B000} are the entropy densities of a HRG with hadron spectra in our quark model.
They are compared with the lattice data, shown in the blue band.
An entropy density is a good measure for the abundance of thermally excited hadrons. 
Regarding a typical hadron volume to be $\sim 1\,{\rm fm}^{-3}$, thermally overlapped hadrons 
are supposed to carry the entropy density of $s \sim 2$-$3\,{\rm fm}^{-3}$ (green band), 
and it gives a rough estimate of the phase transition temperature from a HRG to a QGP. 
At low temperature and $B=0$, pions are dominant, but for $T \gtrsim 100$ MeV other massive excitations make considerable contributions.

One of serious drawbacks from the use of the purely non-relativistic expression is that the entropy density at $T \lesssim 100$ MeV is too large.
This must be related to pions.
Indeed, 
for $K \lesssim M \sim 600-800$ MeV, a non-relativistic kinetic energy is $\vK^2/2M$ (thin red line), smaller than $\sqrt{m_\pi^2+\vK^2} - m_\pi$.
Hence, the non-relativistic spectra at finite $K$ lead to too many thermally excited pions and overpredict the entropy.
For a HRG at $B=0$, this artifact is largely cured by replacement ($E_{\rm NRQ}$ in Eq.(\ref{eq:E_NRQ_B0}))
\beq
E_{\rm NRQ} (\vK) ~ \rightarrow ~ E_{\rm R} = \sqrt{ E_{\rm in}^2 + \vK^2 } \,,
\eeq
with which the entropy density (bold red line in Fig.\ref{fig:total_s_Bdep_B000}) at low $T\simeq 110$ MeV  becomes consistent with the lattice data.

After the relativistic replacement, we still observe that the entropy density of our quark model is still larger than in the lattice.
It turns out that non-relativistic quark models overpredict the excited states at energies greater than $\sim 0.9$ GeV. 
In Fig.\ref{fig:B0_histgram}, we show the histogram for the number of states for mesonic spectra in our quark model, 
and for mesonic and baryonic spectra from the list of the Particle Data Group (PDG)  \cite{ParticleDataGroup:2020ssz}.
The overpredicted spectra at $E \ge 0.9$ GeV affect the entropy density around $T\gtrsim 110$ MeV.
This trend will be also seen at finite $B$ in the next section.

\subsection{At finite $B$}
\label{sec:HRG_finite_B}

\begin{figure}
\resizebox{0.5\textwidth}{!}{%
  \includegraphics{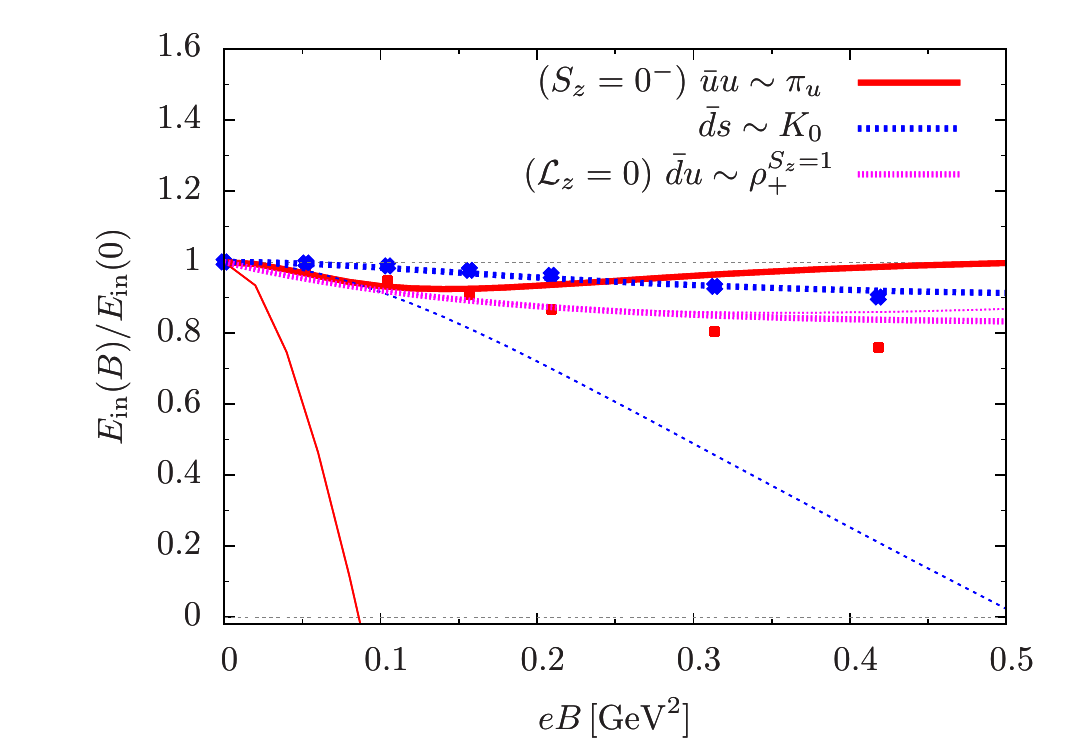}
}
\caption{ 
The ratio $E_{\rm in}(B)/E_{\rm in}(0)$ for neutral and charged mesons.
For neutral mesons we show the results for $\pi_u$ and $K_0$, together with the lattice data in Ref.\cite{Ding:2020hxw}.
For a charged meson we show the result for $\rho_+^{S_z=1}$ with $\calL_z=0$.
The thin lines are the results for a fixed $\alpha_s$.
}
\label{fig:mass_fixed_alpha}       
\end{figure}

\begin{figure*}

\vspace{-0.3cm}
\begin{center}
\resizebox{0.9\textwidth}{!}{
  \includegraphics{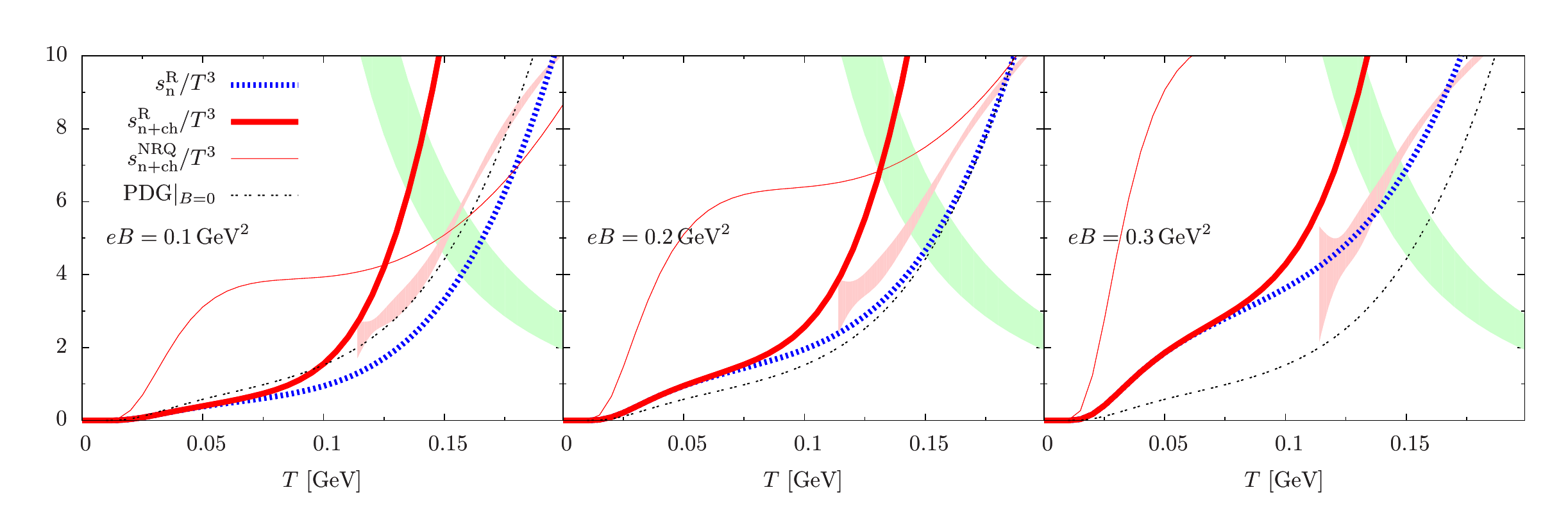}
}
\end{center}
\vspace{-0.3cm}
\caption{Normalized entropy densities at $eB=0.1, 0.2,$ and $0.3\, {\rm GeV}^2$ as functions of $T$. 
We plot the results for neutral mesons ($s_{\rm n}^{\rm R}$), neutral plus charged mesons ($s_{\rm n+ch}^{\rm R}$) with relativistic corrections, and $s_{\rm n+ch}^{\rm NRQ}$ within pure non-relativistic treatments.
As a guideline we also plot the HRG result at $B=0$ which is based on the PDG list for mesonic and baryonic spectra.
The lattice results of Ref.\cite{Bali:2014kia} are shown in the red band.
The constant entropy density $s=2-3\,{\rm fm}^{-3}$ is also shown in the green band.
}
\label{fig:entropy_at_finite_Bs}       
\end{figure*}

As discussed for the $B=0$ case,
the HRG results depend on whether we treat the center of mass motion in a relativistic way or not.
Unlike the $B=0$ case, it is not straightforward to find a proper expression for the relativistic energy.
Hence the following treatments should be regarded as phenomenological.

\subsubsection{Neutral mesons}
\label{sec:HRG_neutral_mesons}

With this precaution, we first consider neutral mesons.
Starting with our nonrelativistic spectrum (see Eqs.(\ref{eq:E_K}), (\ref{eq:E_in}), and (\ref{eq:H_spin})),
\beq
E^{\rm NRQ}_{ {\rm n_i} } (\vK)
= \frac{\, K_z^2 + (1-\eta) \vK_\perp^2 \,}{2M} + E_{\rm in} \,,
\eeq
we infer the relativistic form as
\beq
E^{\rm R}_{ {\rm n_i} } (\vK)
\equiv \sqrt{ K_z^2 + (1-\eta) \vK_\perp^2 + E_{\rm in}^2 \,} \,.
\eeq
As in the $B=0$ case, this phenomenological modification reduces thermal contributions from low-lying mesons.

With some qualification on the flavor multiplet (discussed below), 
the pressure from the $i$-th flavor neutral meson is given by 
($E_{\rm n_i}$ is either $E^{\rm NR}_{\rm n_i}$ or $E^{\rm R}_{\rm n_i}$)
\beq
&&\!\!\! P^{\rm th}_{\rm n_i} 
= -T \! \!\sum_{n_\perp,\, l,\, n_z} \sum_{ {\rm spins} } \int_{\vK} \ln \big(\, 1- \rme^{ - E_{\rm n_i} (\vK) /T } \, \big)
\nonumber \\
&&\!\! =- \frac{T}{\, 1-\eta \,} \! \sum_{n_\perp,\, l,\, n_z} \sum_{ {\rm spins} } \int_{\tilde{\vK} } \ln \big(\, 1- \rme^{ - E_{\rm n_i} ( \tilde{\vK} ) /T } \, \big) \,,
\eeq
where $\int_{\vK} = \int \frac{\, \rmd \vK \,}{\, (2\pi)^3 \,}$ and we have rescaled the integration variables, $\vK \rightarrow \tilde{\vK} = ( K_z, \sqrt{1-\eta \,} \, \vK_\perp )$. 
The total pressure is given by $P_{\rm n}^{\rm th} = \sum_{i \in {\rm flavor} } P_{\rm n_i}^{\rm th} $.
This expression is used to evaluate the entropy density $s_{\rm n} = \partial P^{\rm th}_{\rm n}/\partial T$.

There is one qualification when we sum up neutral mesons in the $S_z=0^-$  channel (which become the pseudoscalar channel at $B=0$). 
For this channel we assume the flavor eigenstates to be $( \bar{u}u - \bar{d}d )/\sqrt{2}$, $( \bar{u}u + \bar{d}d -2\bar{s}s )/\sqrt{6}$, 
and drop off the contribution from the $SU(3)$ singlet, $( \bar{u}u + \bar{d}d + \bar{s}s )/\sqrt{3}$. 
If we do not organize states in this way there would be two light mesons ($\bar{u}u$ and $ \bar{d}d )$ and one heavy boson ($ \bar{s}s$); 
this should be artifacts of neglecting the $q\bar{q}$ annihilations and the topological susceptibility which lift up the flavor singlet mass. 
Meanwhile, for the other channels we do not apply such arrangement in flavors and directly use the spectra of $\bar{l}l$ $(l=u,d,s)$. 
This treatment is consistent with the mass splitting $m_\rho \simeq m_\omega < m_\phi$.

\subsubsection{Charged mesons}
\label{sec:HRG_charged_mesons}

As in the neutral meson case, we infer the relativistic form for the center of mass energy.
With our non-relativistic spectrum
\beq
E^{\rm NRQ}_{ {\rm ch_i} } (K_z)
= \frac{\, K_z^2  \,}{2M} + E_{\rm in} \,,
\eeq
we infer the relativistic form as
\beq
E^{\rm R}_{ {\rm ch_i} } (K_z)
\equiv \sqrt{ K_z^2  + E_{\rm in}^2 \,} \,.
\eeq
We are less sure about the validity of the expression than in the neutral meson case;
here the center of mass motion and the relative motion couple and they are encoded into $E_{\rm in}$.
Meanwhile, the ground state spectrum at finite $B\gtrsim \lqcd^2$ tends to appear at higher energy ($\gtrsim 500 $ MeV) than in the neutral meson case ($\sim 100$ MeV),
so the artifacts are expected to appear at $T \gtrsim 100$ MeV.

The contribution from a particular flavor state is
($E_{\rm ch_i}$ is either $E^{\rm NRQ}_{\rm ch_i}$ or $E^{\rm R}_{\rm ch_i}$)
\beq
\!\!\! P^{\rm th}_{\rm ch_i} 
&=& -T  \frac{\, |B_R| \,}{\, 2\pi \,}  \sum_{ \calL_z = -\infty}^{ \infty } \sum_{ {E_0^\perp} {\rm levels} }^\infty \sum_{n_z=0}^\infty \sum_{ {\rm spins} } \nonumber \\
&& \times \int_{-\infty}^{\infty} \frac{\, \rmd K_z \,}{\, 2\pi \,} 
	 \ln\big(\, 1- \rme^{ - E_{\rm ch_i} /T } \, \big) \,.
\eeq
The factor $|B_R|/2\pi  $ comes from the summation of $N_\calK$, see Sec.\ref{sec:N_calK} for the derivation.
The total pressure is obtained from the sum over all flavor multiplets, $P_{\rm ch}^{\rm th} = \sum_{i \in {\rm flavor} } P_{\rm ch_i}^{\rm th} $.
The entropy is given by $s_{\rm ch} = \partial P^{\rm th}_{\rm ch}/\partial T$.

\subsection{HRG: numerical results at finite $B$ }
\label{sec:HRG_numerical}

For comparisons of our HRG with the lattice results, we begin with the low-lying spectra for which lattice results are available.
In Fig.\ref{fig:mass_fixed_alpha} we show 
the ratio $E_{\rm in}(B)/E_{\rm in}(B=0)$ for charge neutral $\bar{u}u$ and $\bar{d}s$ mesons with $S_z=0^-$,
and a charged $\bar{d}u$ state with $S_z=1$ and $\calL_z=0$.
The lattice data in Ref.\cite{Ding:2020hxw} are shown for the $\bar{u}u$ and $\bar{d}s$ mesons.

Our quark model results with the running coupling seem reasonably consistent with the lattice results for $|eB| \lesssim 0.15\, {\rm GeV}^2$ in Fig.\ref{fig:mass_fixed_alpha}.
At larger $B$, the state $\bar{u}u$ begins to slightly deviate from the lattice data while our result for $\bar{d}s$ remains consistent with the data. 
But it should be kept in mind that the $B$-dependence of the spectra is sensitive to our treatments of the short range correlations,
as one can see from the results for a fixed $\alpha_s$ where some modes become unstable at large $B$.
The QCD running coupling tempers the short range correlations at large $B$.

Now, with reasonable descriptions of low-lying meson masses at $|eB| \lesssim 0.15\, {\rm GeV}^2$, 
we examine the low temperature thermodynamics.
Shown in Fig.\ref{fig:entropy_at_finite_Bs} are the entropy densities of a neutral meson gas for various $B$ and $T$. 
We plot the results for the neutral mesons ($s_{\rm n}^{\rm R}$), neutral plus charged mesons ($s_{\rm n+ch}^{\rm R}$) with relativistic corrections, 
and $s_{\rm n+ch}^{\rm NRQ}$ within pure non-relativistic treatments.
They are compared with the lattice results in Ref.\cite{Bali:2014kia}.
As a guideline we also plot the HRG result at $B=0$ which is based on the PDG list for mesonic and baryonic 
spectra\footnote{In 
Ref.\cite{Endrodi:2013cs}, the author computed the PDG based HRG entropy at finite $B$, regarding hadrons as elementary particles.
The resulting entropy density to $T \sim 100$ MeV is found to be very close to the $B=0$.
}.
As we have mentioned before, our HRG includes only mesons and the resulting entropy should be smaller than in the lattice.
The baryon masses are $\gtrsim 0.9$ GeV, so we expect the corrections become substantial for $T\gtrsim 100$ MeV.

The most important consequence of magnetic fields is that they increase the phase space for neutral mesons; 
at large $B$, the phase space enhancement of a factor $\sim (1-\eta)^{-1} \sim |B|^2/\lqcd^4$ takes place.
At low temperature where the lightest neutral mesons dominate, 
the entropy density is significantly larger at finite $B$ than the $B=0$ case. 
This tendency is very different from the PDG based HRG at finite $B$, where neutral states are treated as elementary and do not depend on $B$;
the resulting entropy density is much smaller than ours and lattice results for $eB\gtrsim 0.1\,{\rm GeV}^2$, see Fig.10 in Ref.\cite{Endrodi:2013cs}.

As found in the $B=0$ case, our model predicts the entropy densities larger than in the lattice.
At $B=0$ we found too many states for $E \gtrsim 0.9$ GeV, and we expect the same situation at finite $B$.
We suspect that the validity of our HRG is limited to $T\sim 100$ MeV at $eB\sim 0.1\, {\rm GeV}^2$, 
and the domain of the validity shrinks as $B$ increases,
as some of overpopulated spectra intrude into the low energy domain.

\section{Discussions}
\label{sec:discussions}

During the analyses of meson spectra and the resulting HRG, several problems were found in the direct application of the conventional non-relativistic quark models.
Here we summarize the problems and discuss possible resolutions:

(i) At large $B$, the short range potentials should be suitably extended to cover the dynamics from the scale $\sim \lqcd$ to $\sim B^{1/2}$.
For meson spectra at finite $B$, it is important to take into account the running of $\alpha_s$.
In this work we tried only the simplest one-loop perturbative expression for $\alpha_s$,
but its applicability is not obvious as problems in this paper involve momentum transfer of $\lesssim 1$ GeV.
We should go back to the $B=0$ case and examine the running $\alpha_s$ at $\lesssim 1$ GeV in more detail \cite{Deur:2016tte}.
We leave such studies for our future work.

(ii) The relativistic extension of the center of mass energy is found to be crucial for the evaluation of thermodynamic quantities.
At $B=0$ we found that (ad hoc) relativistic extension considerably improves the agreement between our model results and the lattice data.
At finite $B$, however, the relativistic extension is not obvious, 
especially for a charged meson whose center of mass motion and the internal quark dynamics couple in an intricate way.
In this respect our work should be extended to a manifestly Lorentz covariant framework.
The confining potential in the present work should be also improved.

(iii) Our quark model predicts too many states at $E\gtrsim 1$ GeV at $B=0$.
The energy splitting between the low-lying states and excited states should be bigger.
We do not fully understand how to increase the energy splitting, 
but it seems to us that radial excitation energies,
related to our harmonic oscillator potential, are too small. 
Within our non-relativistic model, 
an ad hoc remedy would be to take a stronger harmonic oscillator potential.
But, then, we also need to substantially increase the strength of the color-electric interaction $V_E$ to fit low-lying spectra.
We did not attempt the parameter set leading to $\la V_E \ra \lesssim -400$ MeV,
and within such range the above-mentioned problem was not solved.
Thus we conclude that the problem is intrinsic to our model and cannot be removed by parameter choices.
Fortunately, there are relativistic versions of quark models which reproduce the hadron spectra to $\sim 2$ GeV quite well \cite{Ebert:2009ub}.
After identifying the problems in non-relativistic modeling, we now plan to proceed to the analyses using a relativistic quark model.
We leave the detailed analyses for our future work.

\section{Summary}
\label{sec:summary}

We have studied neutral and charged mesons in magnetic fields. We used a non-relativistic constituent quark model which has been widely used for the hadron spectroscopy; the confinement is implemented through a harmonic oscillator potential, and short range correlations are treated in a perturbative scheme. These schemes are directly used for a system in magnetic fields. Based on the previous works on the quark mass gap \cite{Kojo:2012js,Kojo:2013uua,Kojo:2014gha,Hattori:2015aki}, we assume that the constituent quark masses are $B$-independent. 
Based on these spectra we compute entropy densities within the HRG framework. 
The phase space enhancement of mesons at finite $B$ plays a key role for entropy densities at low $T$.

Through the exercises in this paper we found that the descriptions of short-range correlations, i.e., color-electric and magnetic interactions, are important for hadrons in magnetic fields. The detailed understanding of these interactions is important for the physics of neutron stars in the context of dense QCD \cite{Baym:2017whm,Kojo:2020krb}. 
Near the core of two-solar mass neutron stars quarks should be relativistic and the importance of color-magnetic interactions should be significantly enhanced. 
From this point of view, hadrons in magnetic fields, which can be simulated on the lattice, may be a useful testbed to delineate the properties of short-range correlations \cite{Kojo:2021ugu,Kojo:2021hqh}.

There are obvious things to do for future works. In this work we studied only mesons but it is important to study also baryons to complete the HRG within our model. 
Although baryons have the masses $\gtrsim 1$ GeV, there are large numbers of states that compensate the Boltzmann factor and hence they must be included for $T\gtrsim 100$ MeV. Another subject of interest is to compute the chiral condensates at finite $T$ within the HRG by evaluating the sigma term for each hadron; as we express the hadron spectra in terms of constituent quark masses, we can estimate the sigma term assuming the current quark mass dependence of the constituent quarks \cite{Kunihiro:1990ts}. 
Finally, as discussed in Sec.\ref{sec:discussions}, the relativistic extension of quark models is crucial.
These topics will be discussed elsewhere.

\section*{Acknowledgement}
I would like to thank 
H.-T. Ding for useful discussions on the meson spectra and the lattice data, 
and G. Endr\H{o}di for the lattice data and explanations for it.
This work is supported by NSFC grant No. 11875144.

\appendix

\section{Some calculations}
\label{sec:some_calculations}
\subsection{($\hat{ \vPi}^2, \hat{\vcalK}^2$) in creation and annihilation operators}
\label{sec:creation_annihilation}

When we evaluate 2D vectors such as $\hat{\vr}_\perp$ and $\hat{\vp}_\perp$ operators, 
it is more convenient to work with an algebraic method. We define two sets of the creation-annihilation operators, 
($ \hat{\Pi}_\pm = \hat{\Pi}_x \pm \rmi \hat{\Pi}_y$, $\hat{ \calK }_\pm = \hat{ \calK }_x \pm \rmi \hat{\calK }_y$)
\beq
(\hat{a}^\dag, \hat{a} ) = \frac{1}{\, \sqrt{2|eB| } \,}
\times
 \left\{ \begin{array}{l}
~ (\hat{\Pi}_- \,, \hat{\Pi}_+ ) ~~~~~(eB \ge 0) \\
~ (\hat{\Pi}_+ \,, \hat{\Pi}_- )  ~~~~~(eB < 0) 
\end{array} \right.
\eeq
and
\beq
(\hat{b}^\dag, \hat{b} ) = \frac{1}{\, \sqrt{2|eB| } \,}
\times
 \left\{ \begin{array}{l}
~ (\hat{\calK}_+, \hat{\calK}_-) ~~~~~(eB \ge 0) \\
~ (\hat{\calK}_-, \hat{\calK}_+)  ~~~~~(eB < 0) 
\end{array} \right.
\eeq
where $(\hat{a}, \hat{a}^\dag)$ and $(\hat{b}, \hat{b}^\dag)$ separately satisfy the usual harmonic oscillator algebra,
\beq
\hat{\vPi}^2 = |eB| ( 2 \hat{a}^\dag \hat{a} + 1 ) \,,~~~~ \hat{\vcalK}^2 = |eB| ( 2 \hat{b}^\dag \hat{b} + 1 ) \,,
\eeq
and $\hat{a}^\dag |n_\Pi, n_\calK \ra = \sqrt{n_\Pi+1\,} | n_\Pi+1, n_\calK \ra$, etc.

A few more expressions are used for charged mesons discussed in the main text. 
We note that $\hat{\vp}_\perp = ( \hat{\vcalK} + \hat{\vPi} )/2$ and $e \vB \times \hat{\vr}  =  \hat{\vcalK}  - \hat{\vPi}$.
Finally, using Eq.(\ref{eq:expand_Pi_calK}),
\beq
2 e\vB \cdot \vl =  \hat{ \vPi }^2 - \hat{ \vcalK }^2  = 2 |eB| \big( \hat{b}^\dag \hat{b} - \hat{a}^\dag \hat{a} \big) \,.
\eeq
%

%

\subsection{Rearrangement of $\vPi_r$ }
\label{sec:rearrangement}

To compute Eqs.(\ref{eq:tPi_r^2}) and (\ref{eq:pi_R_pi_r}), we note 
\beq
\hat{ \tilde{\vPi} }_r = \hat{ \vp }_r  - \frac{1}{\, 2 \,} \vec{\calB} \times \hat{ \vr } 
 = \hat{ \vPi }_r - \frac{1}{\, 2 \,} ( \vec{\calB} - \vB ) \times \hat{ \vr }
 \,,
\eeq
and $\hat{ \tilde{\vcalK} }_r = \hat{ \tilde{\vPi} }_r + \vec{\calB} \times \hat{\vr}$. Eliminating $\vec{\calB} \times \hat{\vr}$,
\beq
\hat{ \vPi }_r 
&=& \frac{1}{\, 2 \,} \bigg(1 + \frac{\, B \,}{\calB} \bigg) \hat{ \tilde{\vPi} }_r + \frac{1}{\, 2 \,}  \bigg(1 - \frac{\, B \,}{\calB} \bigg) \hat{ \tilde{\vcalK} }_r  
\nonumber \\
&\equiv & f_+ \hat{ \tilde{\vPi} }_r  + f_-  \hat{ \tilde{\vcalK} }_r\,.
\eeq
%
The last expression will be used when we evaluate the coupling $\hat{ \vPi }_R \cdot \hat{ \vPi }_r$ which will appear in computations of charged mesons.

\subsection{$N_\calK$ and the density of states}
\label{sec:N_calK}

We have not discussed any constraints on $N_\calK$ (except $N_\calK \ge 0$), which would give an impression that $N_\calK$ has no upper bound. At this stage we have to be careful about the counting of the density of states (for the detailed discussions, e.g. Ref.\cite{Hattori:2015aki}). For this purpose we consider the system size of $V_2 = \pi R^2$. The momenta $\calK_R$ characterizes the guiding center of the cyclotron orbit measured from the origin, and its radius is $|\vcalK_R/B_R | = \sqrt{2 N_\calK/|B_R|}$ which must be smaller than $R$. 
Thus the maximum of $N_\calK$ for a given volume $V_2$ is $N_\calK^{\rm max} = R^2 |B_R|/2 = V_2 \times | B_R |/2\pi$. Taking this into account, the sum of states per volume  is
%
\beq
\frac{1}{\,V_2 \,} \sum_{ N_\calK = 0}^{N_\calK^{\rm max} }  
= \frac{\, |B_R| \,}{\, 2\pi \,} \,. 
\eeq
%


\end{document}